\documentclass[useAMS,usenatbib,A4]{mn2e}
\usepackage{graphicx,natbib,times,deluxetable}
\title[Galaxy Zoo: Blue Early-type Galaxies]{Galaxy Zoo: A
  sample of blue early-type galaxies at low redshift\thanks{This publication has been made possible by the participation of more than 160,000 volunteers in the Galaxy Zoo project. Their contributions are individually acknowledged at
\texttt{http://www.galaxyzoo.org/Volunteers.aspx}.}}
\author[Kevin Schawinski et al.]{
  \parbox[t]{16cm}{
  Kevin Schawinski$^{1,2,3}$\thanks{E-mail: kevins@astro.ox.ac.uk},
  Chris Lintott$^{3}$,
  Daniel Thomas$^{4}$,
  Marc Sarzi$^{5}$,
  Dan Andreescu$^{6}$,
  Steven P. Bamford$^{4,7}$,
  Sugata Kaviraj$^{3}$, 
  Sadegh Khochfar$^{3,8}$,
  Kate Land$^{3}$,
  Phil Murray$^{9}$,
  Robert C. Nichol$^{4}$,
  M. Jordan Raddick$^{10}$,
  An\v{z}e Slosar$^{11}$,
  Alex Szalay$^{10}$,
  Jan VandenBerg$^{10}$ and
  Sukyoung K. Yi$^{12}$\\
  }\\
$^{1}$Department of Physics, Yale University, New Haven, CT 06511, U.S.A.\\
$^{2}$Yale Center for Astronomy and Astrophysics, Yale University, P.O. Box 208121, New Haven, CT 06520, U.S.A.\\
$^{3}$Department of Physics, University of Oxford, Oxford OX1 3RH, UK.\\
$^{4}$Institute of Cosmology \& Gravitation, University of Portsmouth, Portsmouth, PO1 2EG.\\
$^{5}$Centre for Astrophysics Research, University of Hertfordshire, College Lane, Hatfield, Herts AL10 9AB.\\
$^{6}$LinkLab, 4506 Graystone Ave., Bronx, NY 10471, USA.\\
$^{7}$Centre for Astronomy and Particle Theory, University of Nottingham,  University Park, 
Nottingham, NG7 2RD, UK.\\
$^{8}$Max Planck Institut f\"{u}r extraterrestrische Physik, P.O. box 1312, D-85478 Garching, Germany\\
$^{9}$Fingerprint Digital Media, 9 Victoria Close, Newtownards, Co. Down, Northern Ireland, BT23 7GY, UK.\\
$^{10}$Department of Physics and Astronomy, The Johns Hopkins University, Baltimore, MD 21218.\\
$^{11}$Berkeley Center for Cosmological Physics, Lawrence Berkeley National Lab, 1 Cyclotron Road, MS 50-5005, Berkeley, CA 94720.\\
$^{12}$Department of Astronomy, Yonsei University, Seoul 120-749, Korea.\\
}

\begin{document}

\newcommand\aaps{{AAPS}}
\newcommand\aj{{AJ}}
\newcommand\araa{{ARA\&A}}
\newcommand\apj{{ApJ}}
\newcommand\apjl{{ApJ}}
\newcommand\apjs{{ApJS}}
\newcommand\aap{{A\&A}}
\newcommand\nat{{Nature}}
\newcommand\mnras{{MNRAS}}
\newcommand\pasp{{PASP}}

\date{}

\pagerange{\pageref{firstpage}--\pageref{lastpage}} \pubyear{2007}

\maketitle

\label{firstpage}

\begin{abstract}
We report the discovery of a population of nearby, blue early-type
galaxies with high star formation rates ($0.5 < \rm{SFR} < 50~
M_{\odot}yr^{-1}$). They are identified by their visual morphology as
provided by Galaxy Zoo for SDSS DR6 and their $u-r$ colour. We select
a volume-limited sample in the redshift range $0.02 < z < 0.05$,
corresponding to luminosities of approximately $L^{*}$ and above, and
with $u-r$ colours significantly bluer than the red sequence. We
confirm the early-type morphology of the objects in this sample and
investigate their environmental dependence and star formation
properties. Blue early-type galaxies tend to live in lower-density
environments than `normal' red sequence early-types and make up
$5.7\pm0.4\%$ of the low-redshift early-type galaxy population. We
find that such blue early-type galaxies are virtually absent at high
velocity dispersions above $200 \rm ~kms^{-1}$. Our analysis uses
emission line diganostic diagrams and we find that $\sim 25\%$ of them
are actively starforming, while another $\sim 25\%$ host both star
formation and an AGN. Another $\sim 12\%$ are AGN. The remaining
$38\%$ show no strong emission lines. When present and uncontaminated
by an AGN contribution, the star formation is generally intense. We
consider star formation rates derived from H$\alpha$, $u$-band and
infrared luminosities, and radial colour profiles, and conclude that the star formation is
spatially extended. Of those objects that are not currently undergoing
star formation must have ceased doing so recently in order to account
for their blue optical colours. The gas phase metallicity of the
actively starforming blue early-types galaxies is supersolar in all
cases. We discuss the place of these objects in the context of galaxy
formation. A catalogue of all 204 blue early-type galaxies in our
sample, including star formation rates and emission line classification, is provided.
\end{abstract}


\section{Introduction}
\begin{figure*}
\begin{center}

\includegraphics[angle=90, width=0.48\textwidth]{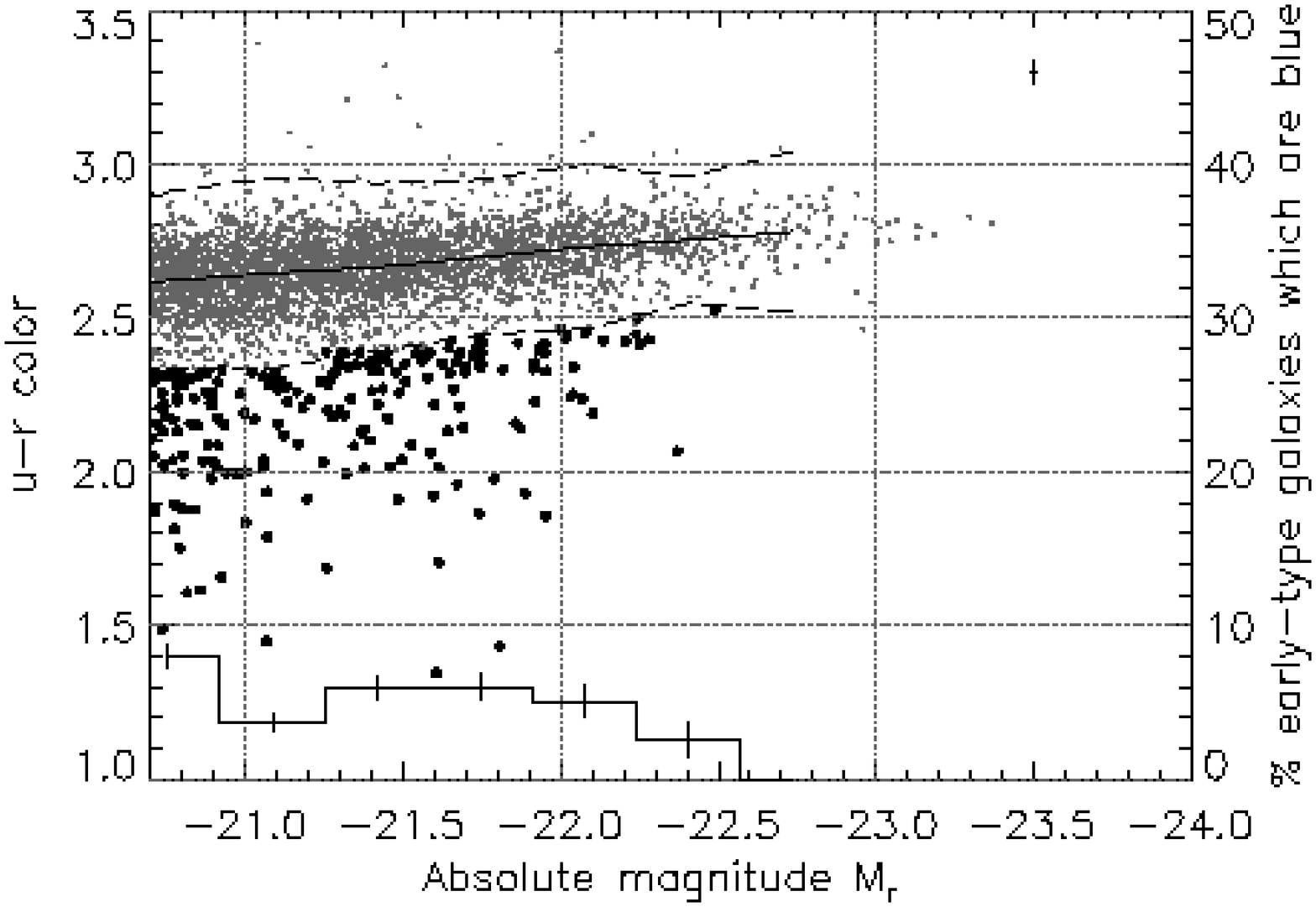}
\includegraphics[angle=90, width=0.48\textwidth]{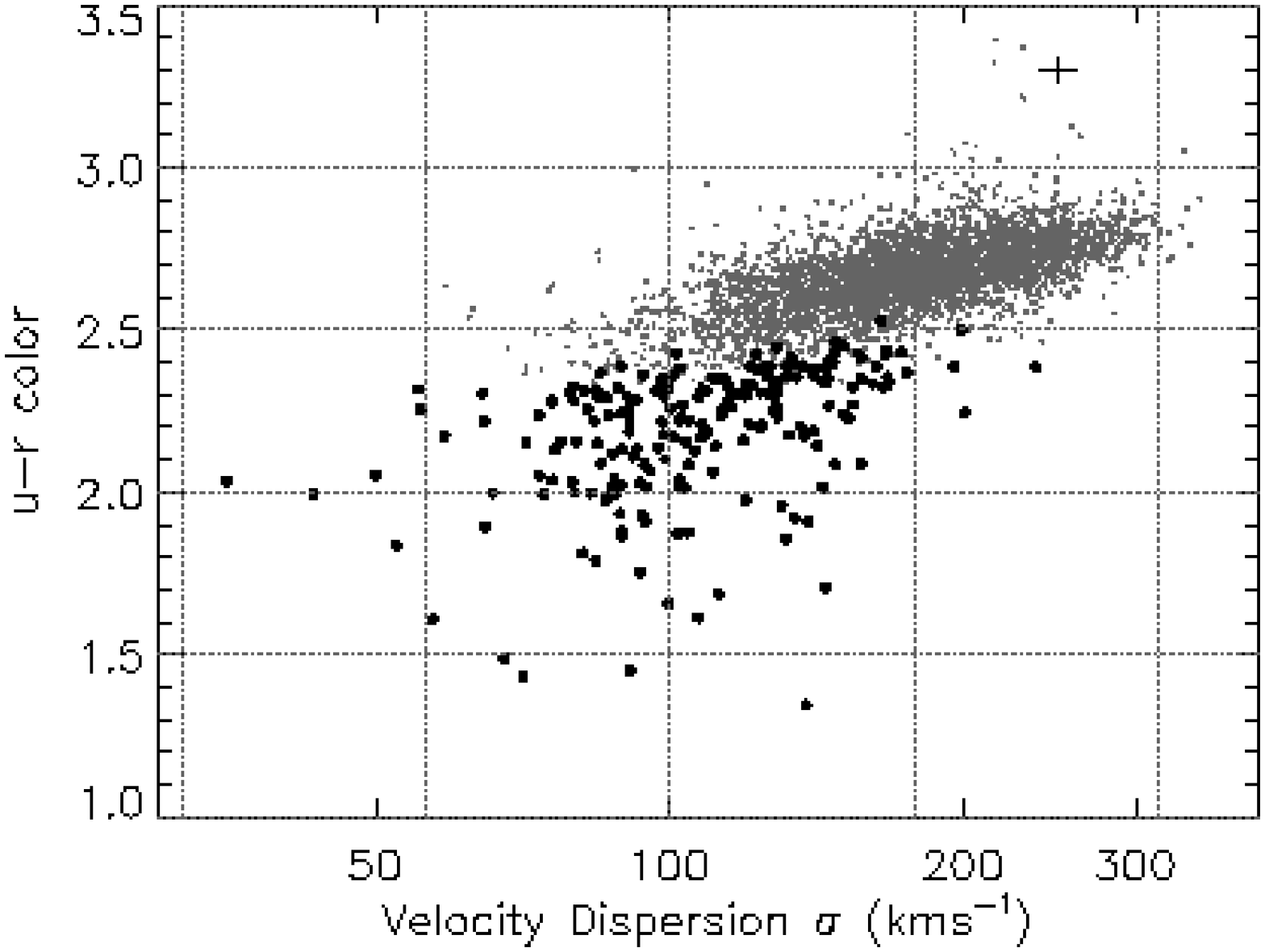}
\caption{\textit{Left:} We show the the optical colour-magnitude
  relation for our volume-limited sample of morphological early-type
  galaxies.  The solid line represents the mean of a gaussian fit to
  the red sequence, while the dashed lines indicate the $3\sigma$
  offset from that mean. We plot all those early-type galaxies above
  this $3\sigma$ offset as small gray points and those below as large
  black points. On top of this, we plot a histogram of the fraction of
  early-type galaxies as a function of $M_{r}$ classified as blue
  ealy-types galaxies and the scale is indicated on the right-hand
  vertical axis. \textit{Right:} We show the colour-$\sigma$ relation for
  our sample, indicating galaxies using the same symbols. Blue
  early-type galaxies have velocity dispersions that are significantly
  lower than the majority of red early-type galaxies with similar
  luminosities. As star formation is suppressed, they will move
  vertically up at a given velocity dispersion as they join the red
  sequence. Blue early-type galaxies in the low-redshift Universe are
  thus building the lower-mass end of the red sequence in accordance
  with downsizing \citep{2005ApJ...621..673T}. The blue early-type
  galaxy fraction is a strong function of velocity dispersion and
  there are \textit{virtually no} blue early-type galaxies above a
  velocity dispersion of $\sigma = 200 \rm ~kms^{-1}$
  (c.f. \citealt{2006Natur.442..888S}).  We indicate the typical
  $3\sigma$ error for the blue early-types in the top right corner.\label{fig:optical_cmr}}
  
  \includegraphics[angle=90, width=\textwidth]{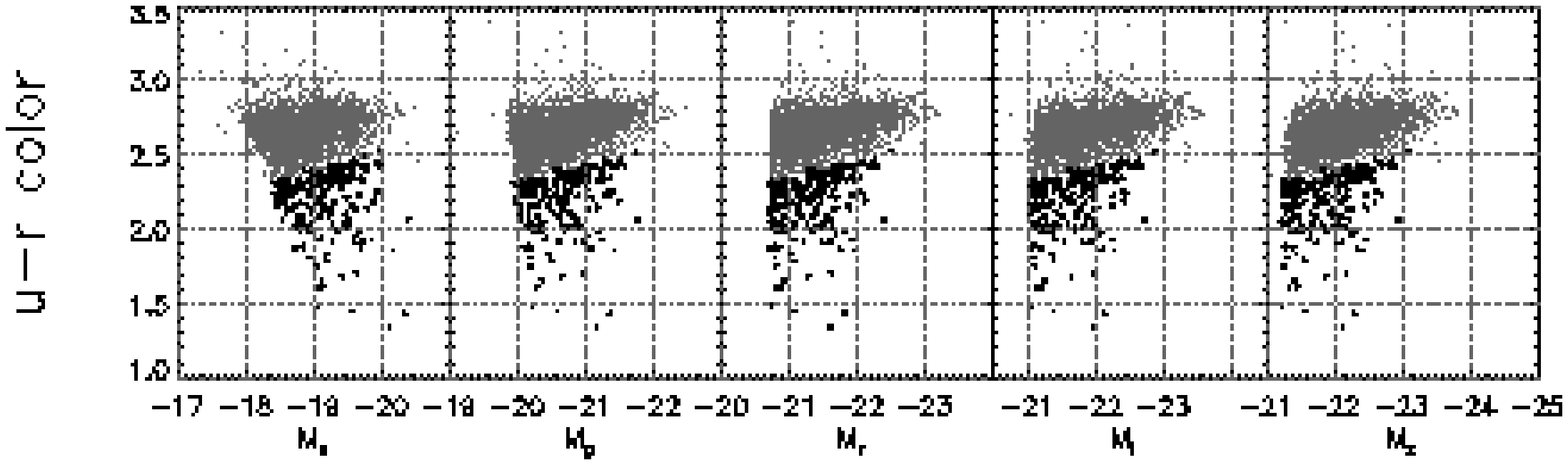}
\caption{We plot the $u-r$ colour-magnitude relations as a function of
all five SDSS filter absolute magnitudes ($M_{u}$, $M_{g}$, $M_{r}$,
$M_{i}$ and $M_{z}$) to illustrate the effect on these absolute
magnitudes of the young stellar populations present in blue early-type
galaxies. In the $M_{u}$ diagram, the young stellar populations boost
the absolute magnitude such that blue early-types are as luminous as
the most massive early-type galaxies despite their low velocity
dispersion (c.f. Figure \ref{fig:optical_cmr}, left). As we go to
redder bands, the absolute magnitude is less and less affected by the
young population and the $M_{z}$ colour-magnitude relation is most
similar to the colour-$\sigma$ diagram.  For an in-depth discussion on
the selection bias introduced by this, see Section
\ref{sec:fj}\label{fig:etype_cmrs}}

\end{center}
\end{figure*}

Early-type galaxies are a fascinating probe of galaxy formation. Their
apparent simplicity is deceiving, since much of the physics that is
involved in their formation and evolution is still poorly understood.
It is now common practice to divide the galaxy population on a
colour-magnitude diagram (e.g. \citealt{2004ApJ...600..681B}; Figure
\ref{fig:optical_cmr}), where early-type galaxies typically appear to
lie on a red sequence while spirals display bluer colours and seem to
reside in a `blue cloud' \citep{1964AJ.....69..635C,
1992MNRAS.254..601B, 2006MNRAS.368..414D,
2007ApJ...665..265F}. Morphological early-type galaxies residing in
this blue cloud of starforming galaxies are of great importance for
our understanding of the formation of early-type galaxies and the
build-up of the red sequence. These blue early-type galaxies
do provide us with a laboratory to understand the physical processes
that underly their evolution \citep{2007MNRAS.382.1415S,
2007ApJS..173..267S, 2008ApJ...673..715C, 2009ApJ...690.1672S}. \cite{2007MNRAS.382.1415S},
hereafter S07, have already shown that there is a significant
population of such blue early-type galaxies at low velocity
dispersion.

Samples of blue early-type galaxies have tended to be small, as the
most reliable way to identify them is through visual inspection of
images. In a seminal study, \cite{1977PhDT.......119H, 1977ApJS...35..171H}
studied the non-Seyfert population of the Markarian catalogue and concludes that
these Markarian systems are systematically blue for their morphology, possibly indicating
an episode of recent star formation. This sample includes a number of systems
classified visually as early-type (E or S0).

Many studies attempt to select large samples by using
automated structural parameter measurements, such as concentration to
isolate early- and late-type galaxies. However, samples selected by
structural quantities do not correspond well to those based on visual
morphological classification \citep{2008ApJ...675L..13V}.
\cite{2004ApJ...601L.127F} found three blue early-type galaxies (two
starforming and one hosting an active-galactic nucleus, AGN) with star
formation rates (SFR) of 2.9 and 4.8 $\rm M_{\odot}yr^{-1}$, an order
of magnitude fainter than the highest SFR we find. A larger sample of
visually classified galaxies is presented in
\cite{2007AJ....134..579F}, though the question of blue early-type
galaxies is not addressed. Early samples of blue galaxies of
early-type morphology tended to focus on very low mass systems,
such as dEs, which may not be the same as the more massive systems that 
we can access in the SDSS universe \citep{1988A&A...204...10K, 1997MNRAS.288...78T, 1997A&AS..124..405D}. A catalogue of early-type galaxies
with emission lines was published by \cite{1987A&AS...67..341B}.

In this Paper we present a new sample of 204 blue early-type galaxies
populating the extreme blue end of a sample of 3588 nearby early-type
galaxies. We thus confirm that blue early-type galaxies with
significant amounts of star formation exist but make up only a small
fraction of early-types in the low-redshift Universe and investigate
their properties in detail. The parent sample is selected by
morphology determined by careful and repeated visual inspection. The
extreme blue sample is then selected based on $u-r$ optical colour.
Our blue early-type galaxies can display intense star-formation, with
star-formation rates ranging from $0.5 < \rm SFR < 50~
M_{\odot}yr^{-1}$, comparable at the low end to typical spiral
galaxies and representing at the high end the most intensely
starforming early-type galaxies ever detected.

\begin{figure*}
\begin{center}

\includegraphics[angle=90, width=0.9\textwidth]{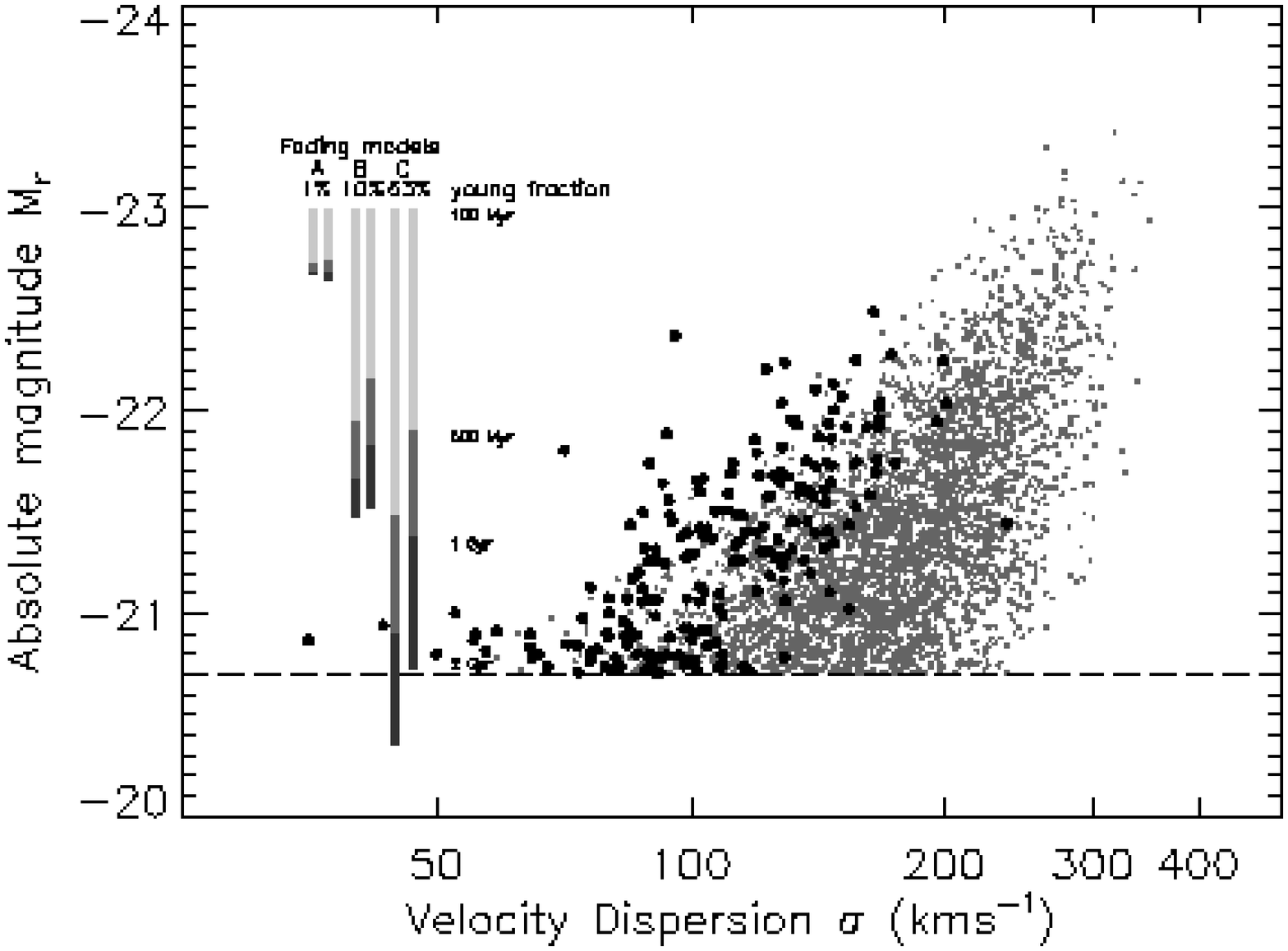}
\caption{In this Figure, we present the Faber-Jackson relation
  \citep{1976ApJ...204..668F} for our volume-limited sample of
  early-type galaxies. The absolute magnitude cut is indicated by the
  dashed line. As in Figure \ref{fig:optical_cmr}, we indicate red
  early-type galaxies as small gray points and blue early-type
  galaxies as larger black points. The blue early-type galaxy
  population scatters off the locus of their red counterparts. This
  may lead to selection bias (see Section \ref{sec:fj}) as the more
  luminous blue early-type galaxies at lower velocity dispersions are
  put into the volume-limited sample. We indicate how blue early-type
  galaxies might fade to the Faber-Jackson relation by determining the
  amount of fading expected from a variety of starburst models. We
  place a young population of 1\%, 10\% and 63\% mass fraction on top
  of an 8 Gyr old solar metallicity SSP with E(B-V) = 0.05
  (\citealt{2000ApJ...533..682C}), corresponding to models A, B and
  C. We then calculate the expected fading starting from 100 Myr age
  to 500 Myr, 1 Gyr and 2 Gyr. We use grayscale to indicate these time
  steps. For each mass fraction, the left bars represent a rapidly
  declining star formation history where the young burst is modelled
  as an exponentially declining star formation history with $\tau =$
  100 Myr. The right bar represents a $\tau =$ 1 Gyr. If these star
  formation histories roughly represent those of blue early-type
  galaxies, then a return to the Faber-Jackson relation within $\sim
  1$ Gyr is plausible for starbursts of a few percent by
  mass.}\label{fig:faber_jackson}

\end{center}
\end{figure*}

\section{Sample Properties}
Our sample of morphological early-type galaxies is drawn from the
Galaxy Zoo \texttt{clean} catalog \citep{2008MNRAS.389.1179L}, which
provides us with visual classifications for galaxies from the Sloan
Digital Sky Survey Data Release 6 (SDSS DR6;
\citealt{2000AJ....120.1579Y, 2008ApJS..175..297A}). The Galaxy Zoo
classifications of morphology are based on the visual inspection of
SDSS images by a large number of independent observers. As described
in \citep{2008MNRAS.389.1179L}, Galaxy Zoo enlists the help of members
of the public to classify the spectroscopic galaxy sample from the
Sloan Digital Sky Survey. This selection made only via morphology
is crucial when selecting non-passive early-type galaxies. We select
all galaxies with spectra in the redshift interval 0.02 $< z <$ 0.05
and create a volume-limited sample by limiting to an absolute
magnitude of $M_{r} < -20.7$, which is slightly below $M_{*}$ for
low-redshift early-type galaxies ($M_{r,*} = -21.15$,
\citealt{2003AJ....125.1849B}). In this volume-limited sample, we then
focus on the morphological early-type galaxies. The visual
classifications from Galaxy Zoo of the objects presented here (see
Section \ref{sec:conf}) should provide robust confirmation of their
early-type morphology as the redshift and apparent magnitude limits
used are lower than those of \cite{2004ApJ...601L.127F} and S07. The colours are derived from SDSS \texttt{modelMags}, which are optimised for galaxy colours, while the absolute magnitudes are based on the SDSS \texttt{petroMag}, which is a good measure of the total light.

\begin{figure*}
\begin{center}

\includegraphics[angle=90, width=0.99\textwidth]{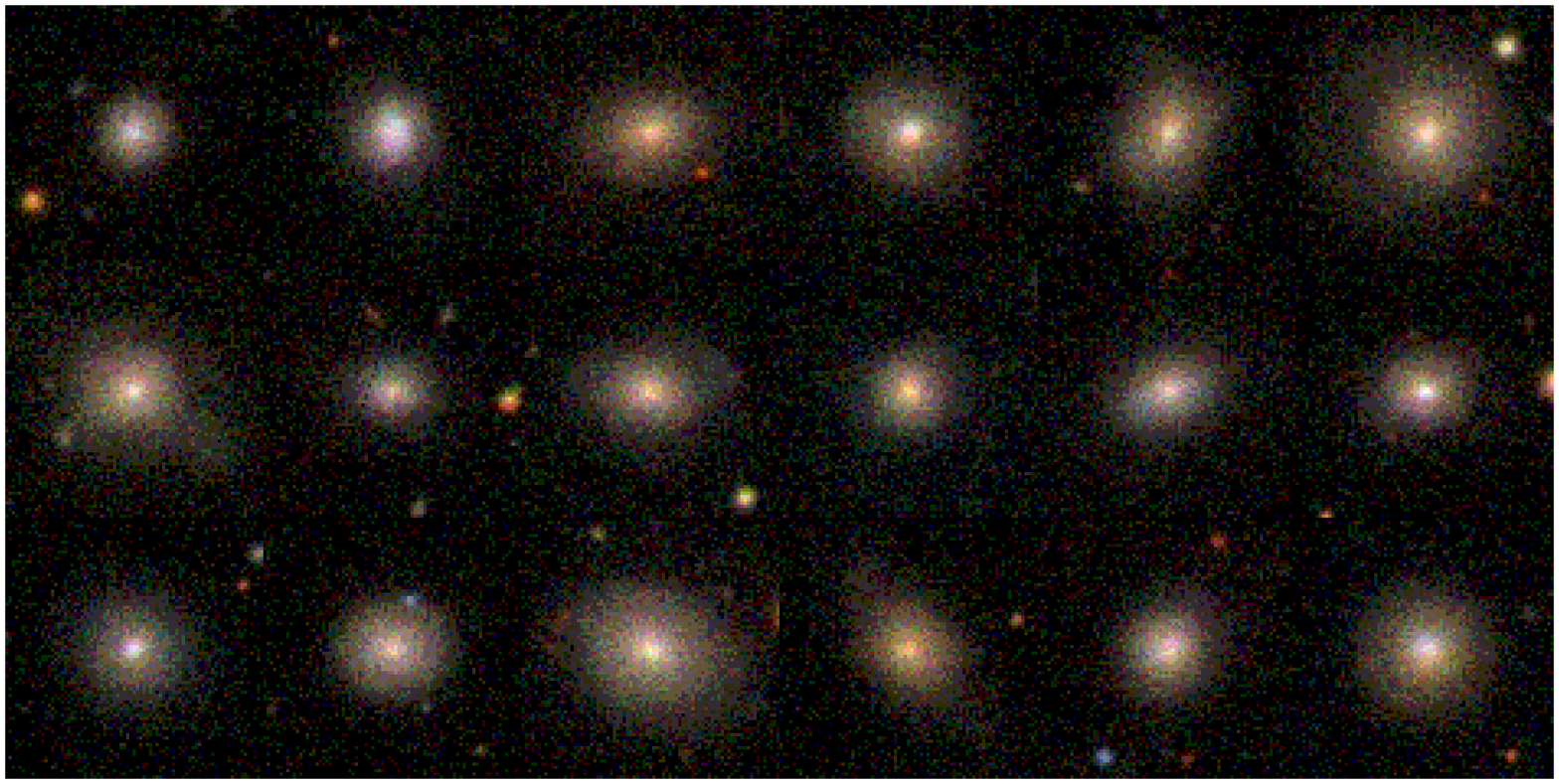}
\caption{In this Figure, we show example SDSS $gri$ composite colour
  images for a sample of blue early-type galaxies. The galaxies shown
  here are all classified as purely starforming (see Section
  \ref{sec:eml}). The objects here all have aperture-corrected
  H$\alpha$ SFR $> 5~ \rm M_{\odot}yr^{-1}$. These objects all have
  higher star formation rates than any previously known
  visually-classified early-type galaxy. The images measure
  $51.2\arcsec \times 51.2\arcsec$.}

\label{fig:example_images}

\includegraphics[angle=90, width=0.99\textwidth]{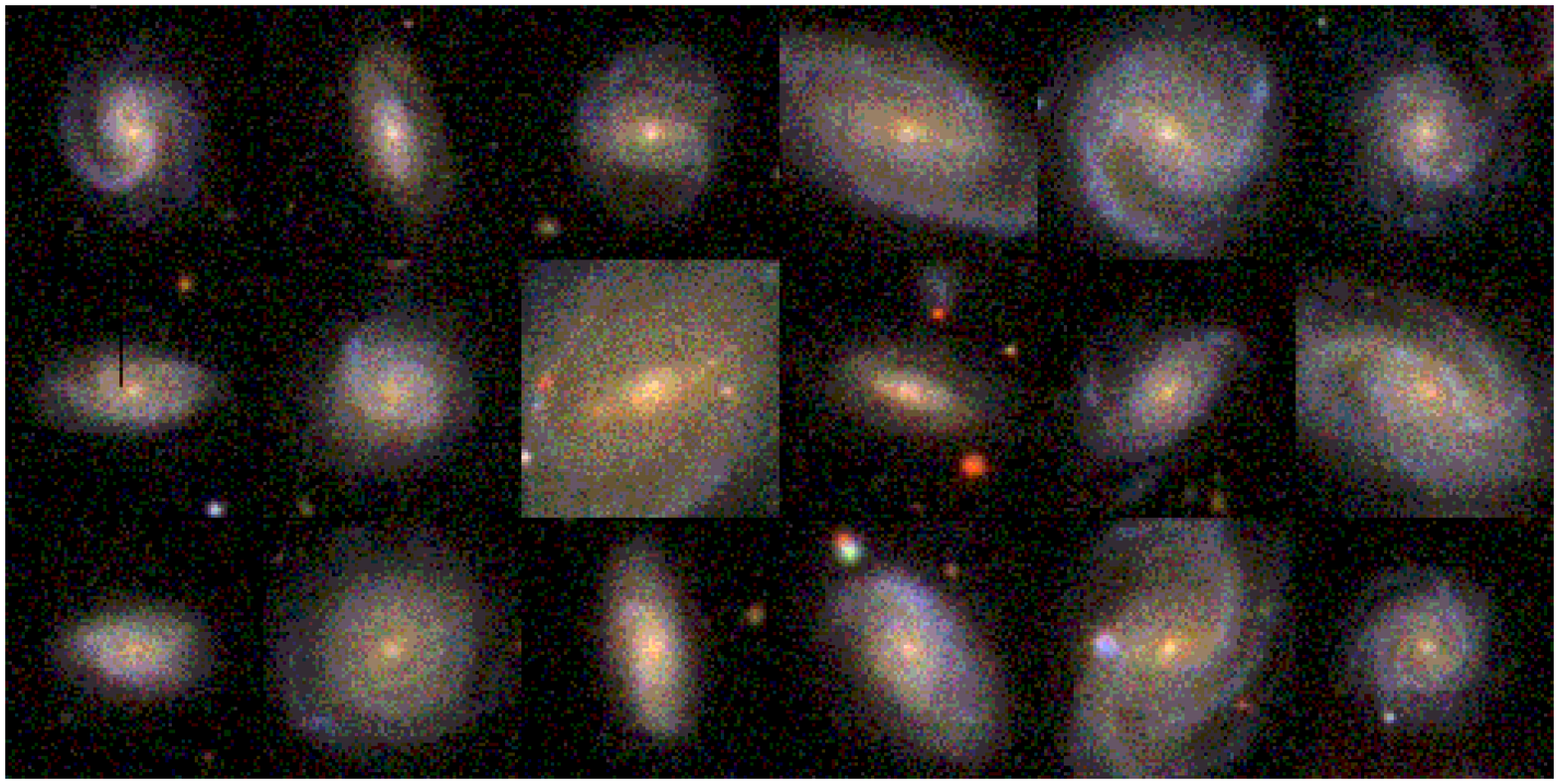}
\caption{In this Figure, we show example SDSS $gri$ composite colour
  images for a sample of face-on spiral galaxies selected by applying
  the same redshift and luminosity cuts that define our blue
  early-type galaxy sample. Their purpose is to indicate the image
  quality in Figure \ref{fig:example_images}, which may not be
  apparent due to the lack of features of early-type galaxies. These
  images also indicate how a starforming disk or spiral arms of any
  prominence appear in images of this quality and so establishes their
  absence in our morphological early-type sample.}

\label{fig:example_spirals}

\end{center}
\end{figure*}

\subsection{Selection of Blue Early-type Galaxies}
\label{sec:selection}
On the left-hand side of Figure \ref{fig:optical_cmr}, we plot the
optical colour-magnitude relation for the early-type galaxies in our
sample. In $u-r$ colour, the majority of early-type galaxies lie on a
tight red sequence, while a minority scatters to significantly bluer
$u-r$ colours. We fit a Gaussian to the $u-r$ colour of the early-type
population in magnitude bins of $dM = 0.33 ~ \rm mag$. Using this fit,
we determine the $3\sigma$ offset to the red sequence and define as
\textbf{blue early-type galaxies} those that are below this
offset. This strong selection ensures that we are probing the blue
extreme of the early-type galaxy population.  On Figure
\ref{fig:optical_cmr}, we indicate the mean of the red sequence as the
solid line and the 3$\sigma$ offsets as dashed lines. This produces a
sample of 204 blue early-type galaxies.

The overall fraction of early-type galaxies that are selected in this
volume-limited sample is $5.7\pm0.4\%$. This fraction is significantly
higher than the 0.23\% expected from a 3$\sigma$ limit of a purely
Gaussian distribution; this is due to the fact that the $u-r$
colour-magnitude relation is not a perfect Gaussian, but displays an
excessive scatter to bluer colours. The assumption of the colour
distribution of the red sequence as a Gaussian is nevertheless a good
one and the blue early-types are a sub-population that scatters off
this Gaussian. The $5.7\pm0.4\%$ fraction is lower than that found by
S07, as our selection via extremely blue colours is more stringent
than their selection via emission lines. Here we only select those
galaxies whose colours are extremely blue, while S07 included
early-type galaxies with emission lines that were closer to, or even
on, the red sequence. Selecting early-type galaxies with emission lines has
a considerable overlap with the colour selection, but there are also numerous
early-type galaxies with emission lines that are on the red sequence. These red
early-type galaxies mostly show emission lines with LINER-like ratios  (see e.g. \citealt{2007ApJ...671..243G} and S07). \cite{2007ApJ...671..243G} show that early-types that show 
LINER emissions tend to be systematically younger than their passive counterparts
at the same velocity dispersion. In this Paper, we are probing the extreme end of
the active early-type population.

If we plot the colour-magnitude relation as a function of all SDSS
bands (see Figure \ref{fig:etype_cmrs}), it becomes apparent that the
young stellar populations that account for the blue $u-r$ colours
affect the absolute magnitude. The $u$-band is most affected,
resulting in absolute magnitudes similar to those of the most massive
red sequence early-type galaxies. The $z$-band is least affected and
is more similar to the colour-$\sigma$ relation. The selection bias
introduced by this are discussed in Section \ref{sec:fj}.

\subsection{The Faber-Jackson Relation and Sample Bias}
\label{sec:fj}
The distributions of $M_{r}$ and $\sigma$ in Figures
\ref{fig:optical_cmr} and \ref{fig:etype_cmrs} support a picture where
blue early-type galaxies are systems containing substantial young
stellar populations with low and intermediate velocity dispersions
almost exclusively below $\sigma = 200 \rm ~kms^{-1}$, in accordance
with the picture of downsizing \citep{2005ApJ...621..673T,
2005ApJ...632..137N, 2005AJ....129...61B}. \textit{There are extremely
few blue early-type galaxies with $\sigma > 200 \rm ~kms^{-1}$ in
the low-redshift Universe.} The absence of blue early-type galaxies at
high velocity dispersions cannot be accounted for by any selection
effects.

In Figure \ref{fig:faber_jackson} we plot the Faber-Jackson relation
\citep{1976ApJ...204..668F}, we can see that blue early-type galaxies
(black points) scatter away from the position of red early-types (gray
points). The velocity dispersions of these systems are unaffected by
young stellar populations, but their boosting of the absolute
magnitude is sufficient to account for the offset. In Figure
\ref{fig:faber_jackson}, we use \cite{2005MNRAS.362..799M} models to
show the expected fading of stellar populations given a selection of
mass fractions and other parameters to show that a declining star
formation history will place blue early-type galaxies on the Faber
Jackson relation within 1 Gyr.

The boosting of the $r$-band luminosity will introduce a bias on our
selection of a volume-limited sample with $M_{r} < -20.7$ as blue
early-type galaxies at lower velocity dispersion will enter this
selection, while red early-type galaxies will not (c.f. Faber Jackson
relation, Figure \ref{fig:faber_jackson}). Thus the fraction of blue
early-type galaxies of $5.7\pm0.4\%$ should be viewed as an upper
limit on the blue fraction given our selection. A sample selection
that is complete in velocity dispersion encounters the problem that
the reliability of the visual inspection decreases as a function of
apparent magnitude \citep{2008arXiv0805.2612B}, and so the purity of
the sample will be compromised. The blue early-type galaxies presented
here are low- to intermediate-mass early-type galaxies, not $M^{*}$
and more massive systems. In a complete, mass-selected sample, the
fraction of blue early-type galaxies at the high-mass end would be
low.

\begin{figure*}
\begin{center}

\includegraphics[width=0.49\textwidth]{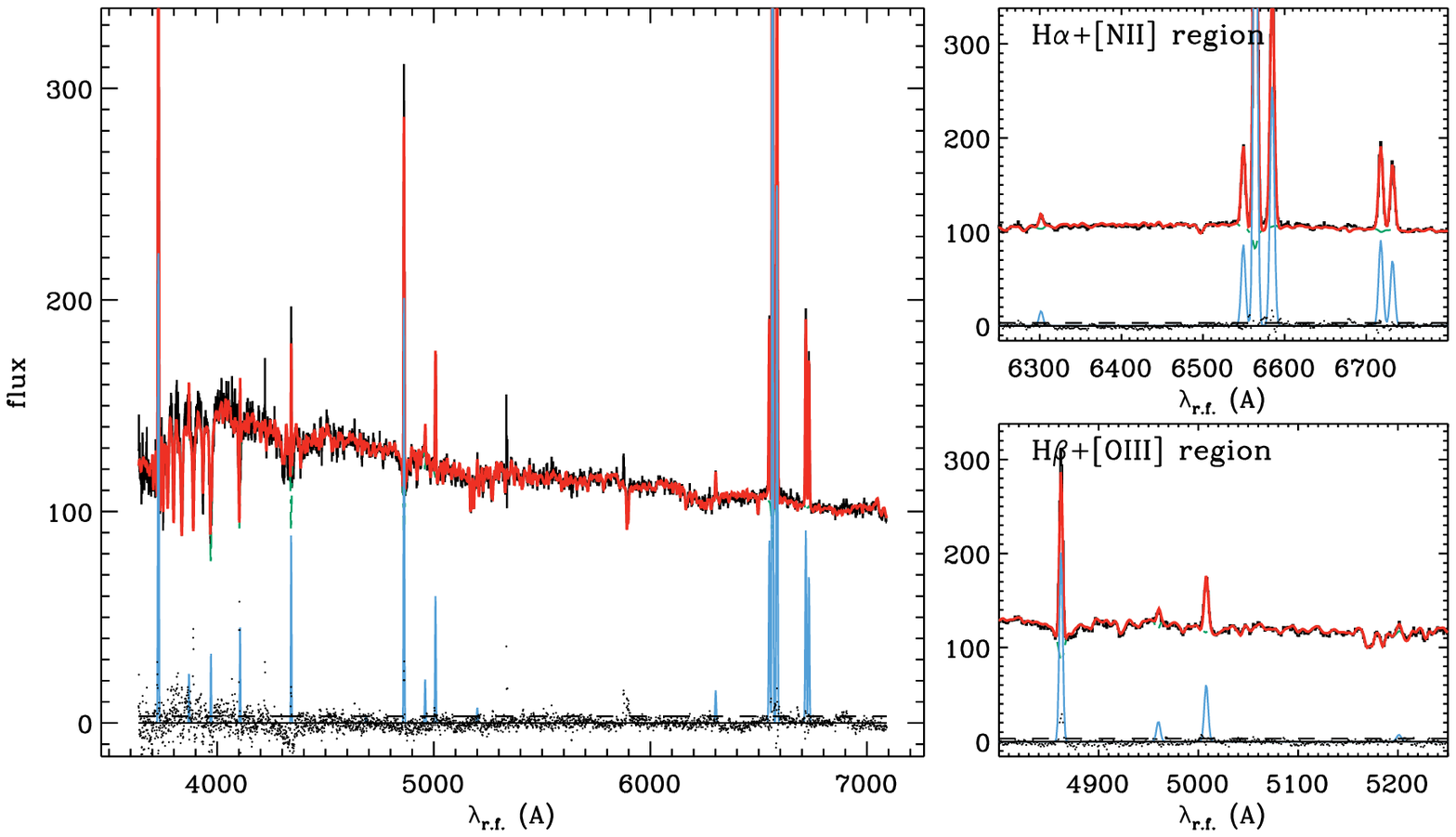}
\includegraphics[width=0.49\textwidth]{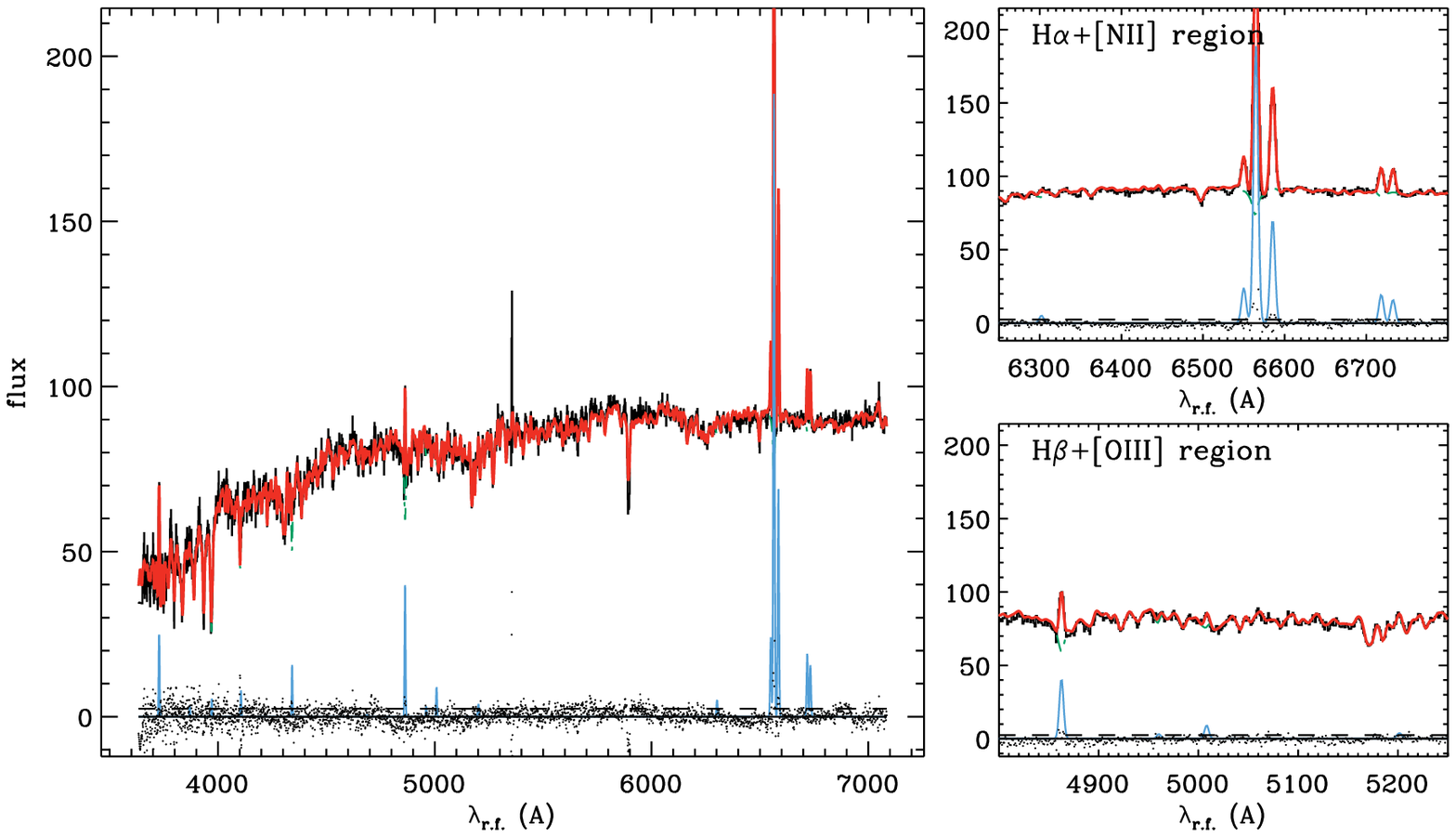}
\includegraphics[width=0.49\textwidth]{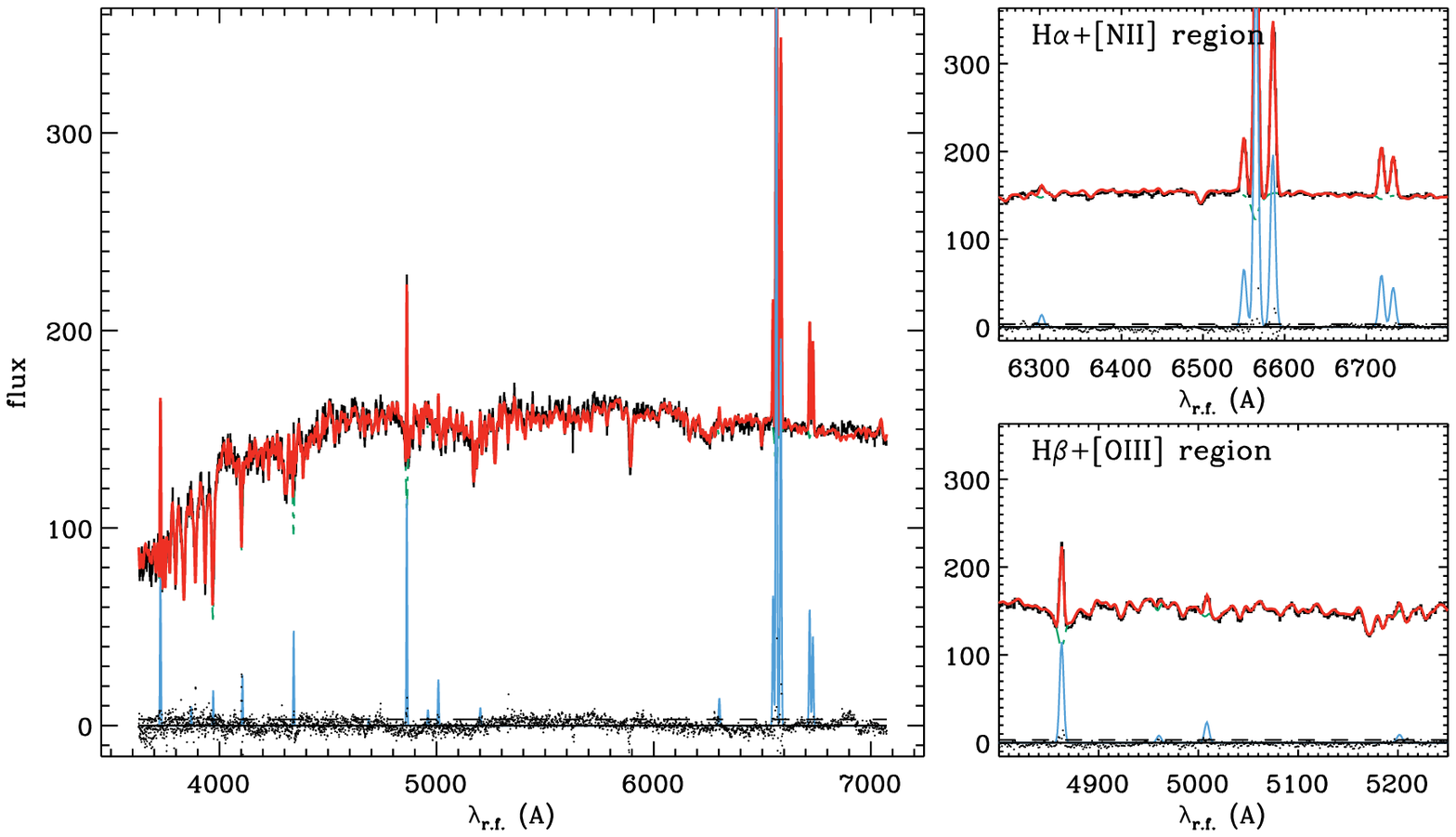}
\includegraphics[width=0.49\textwidth]{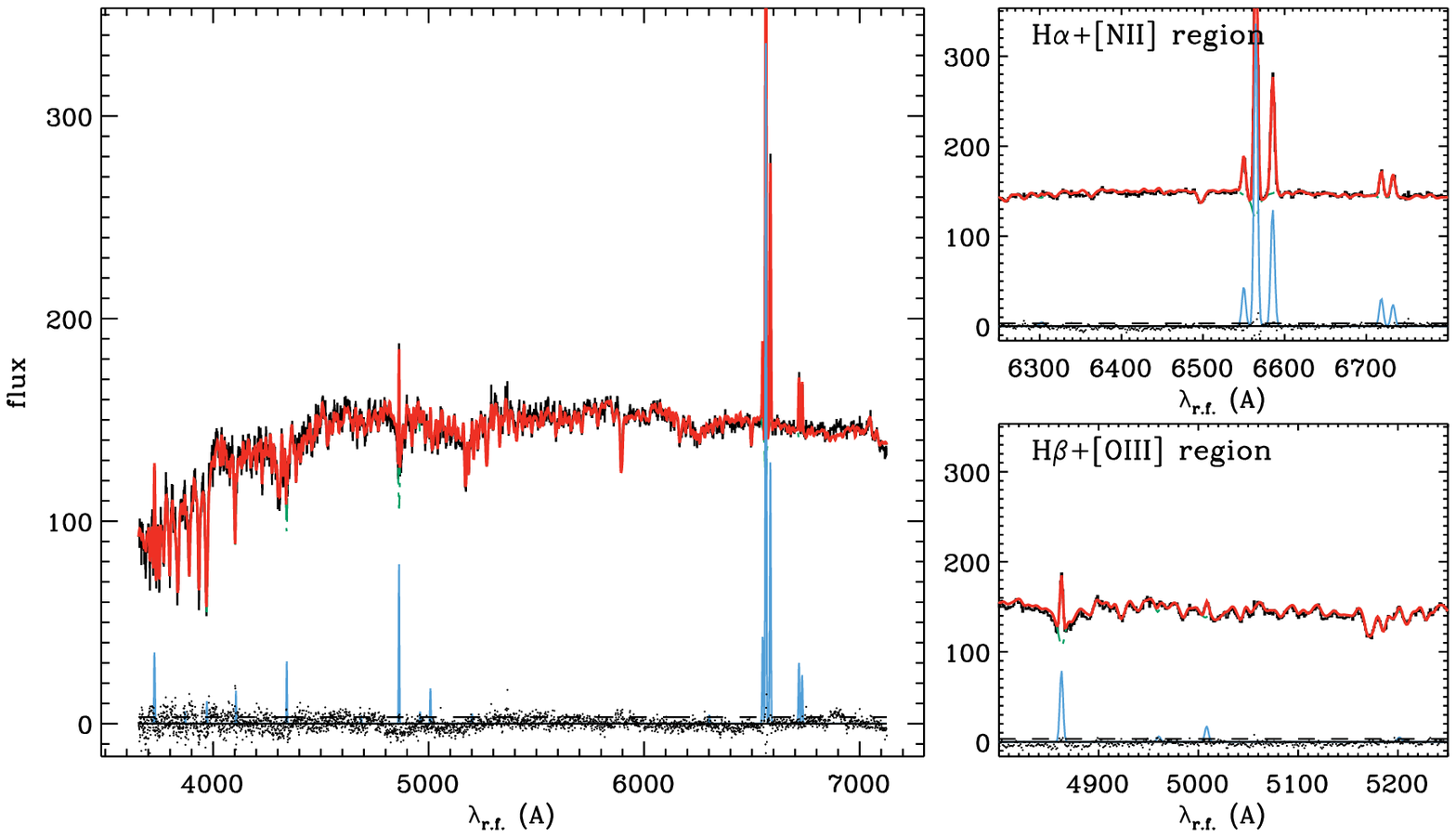}
\caption{We show the SDSS spectra of four blue early-type galaxies
processed with \texttt{GANDALF}. For each galaxy, we show the whole
galaxy spectrum in the restframe. The best fit is overplotted in red,
while the fit to the emission lines is shown in blue. The residuals
for this fit are also shown by the small points at the lower end of
each panel, where the dashed lines further show the average level of
the residuals that is used assess the detection of the emission
lines. We further focus in on two wavelength regions of interest
around H$\alpha$+[NII] and H$\beta$+[OIII]. All four spectra feature
deep Balmer absorption lines and some show considerable dust
extinction.\label{fig:example_spec}}

\includegraphics[angle=90, width=\textwidth]{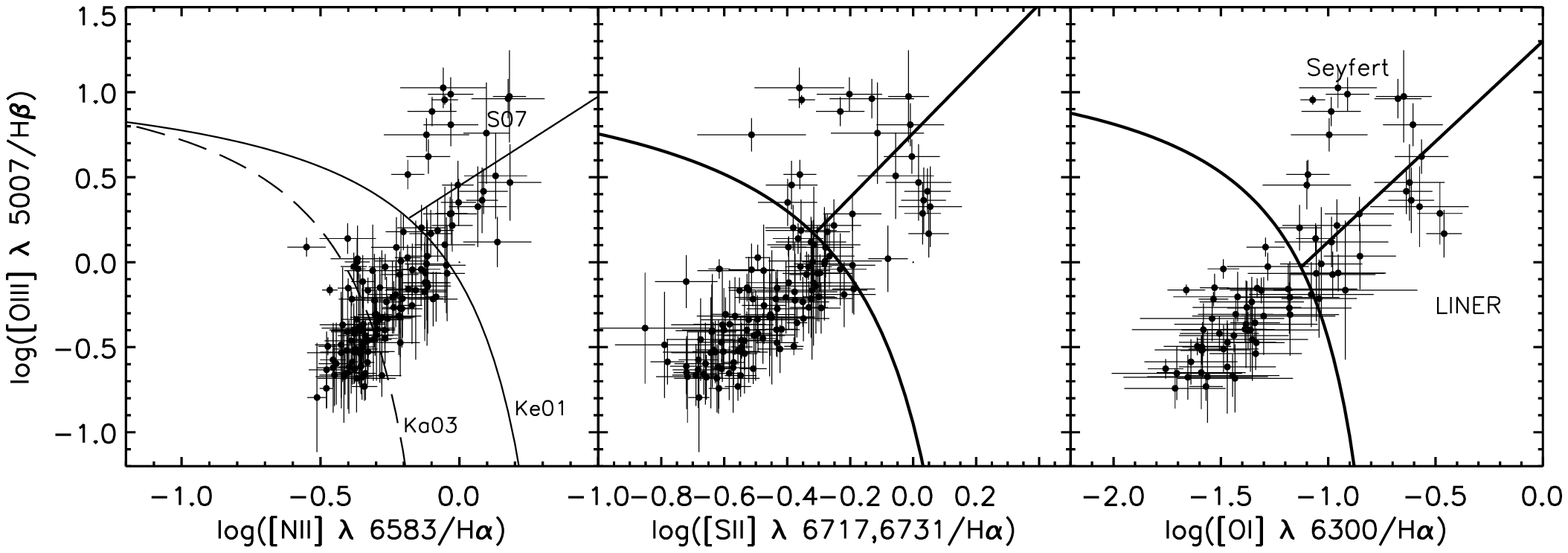}
\caption{In this Figure, we show the emission line ratio diagrams for
our blue early-type galaxy sample (BPT diagram;
\citealt*{1981PASP...93....5B}). Only objects where all four lines in
each diagram are detected with $\rm S/N > 3$ are plotted. In each
panel, we use a different line from left to right: [NII]
$\lambda$6583, [SII] $\lambda$6583 and [OI] $\lambda$6583. All objects
below the line labelled Ka03 \citep{2003MNRAS.346.1055K} are
classified as purely starforming. All objects between the Ka03 line
and the line labelled Ke01 \citep{2001ApJ...556..121K} are AGN+SF
composites. The line from S07 divides Seyferts and LINERS on the [NII]
diagram. All objects beyond this line are separated into Seyfert AGN
and LINERs where possible with the [OI] or [SII]
diagrams.\label{fig:bpt_plots}}

\end{center}
\end{figure*}

\subsection{Confirmation of Early-type Morphology}
\label{sec:conf}
As discussed earlier, this sample of blue early-type galaxies is at
lower redshift and apparent magnitude and so the visual classification
is highly secure. We present example $gri$ images to illustrate this
in Figure \ref{fig:example_images}. We also provide images of a sample
of galaxies classified as face-on spirals in the same luminosity and
redshift range in Figure \ref{fig:example_spirals} for comparison. By
using the Galaxy Zoo \texttt{clean} sample, we further ensure that we
have only selected galaxies about which a substantial majority of
classifiers ($>80\%$ by weighted vote, see
\citealt{2008MNRAS.389.1179L} for a discussion) concur.

The most relevant comparison for Galaxy Zoo early-types is with those of the MOSES (Morphologically Selected Ellipticals in SDSS; \citealt{2007MNRAS.382.1415S}) sample, selected by one professional classifier. Essentially all of the MOSES early-types included in Galaxy Zoo ($>99.9\%$) were found to be early-type. However, the 
Galaxy Zoo early-type sample included a small number of galaxies that were not 
classified as early-type in MOSES; comparison with the sample of \cite{2007AJ....134..579F} who 
provided detailed classifications of $\sim$3000 SDSS galaxies suggests that these extra 
galaxies are likely to be S0 galaxies. Each of the galaxies included in the present 
sample were inspected by an expert classifier (KS) and found to be of early-type 
morphology.

A further advantage of the Galaxy Zoo classifications rests in the fact that we have multiple classifications for every object. This allows us to statistically analyse the distribution of classifications to assess its reliability. In a companion paper, \cite{2008arXiv0805.2612B} statistically assess
the bias of galaxy classifications from Galaxy Zoo and SDSS as
function of luminosity and redshift. This study underlines the need
for our selection of low redshift, high-luminosity galaxies for
obtaining the high reliability of visual inspection needed to rule out
any significant late-type/spiral interlopers in this sample. Based on
the estimates of \cite{2008arXiv0805.2612B}, we conclude that our
sample as a maximum contamination of spiral interlopers of $\sim4\%$,
which corresponds to 8 objects in the blue early-type sample. This
makes our sample exceedingly clean; Sa spiral galaxies with distinct
disks and spiral arms are \textit{not} contributing to our sample. We confirm this by
visually inspecting all 204 objects.

\begin{table}
\begin{center}
\caption{Emission Line Classification Results for blue early-type galaxies}
\label{tab:class_results}
\begin{tabular}{@{}lrrl}
\hline
\hline
Classification & \multicolumn{1}{c}{Number} & \multicolumn{1}{c}{Fraction} \\
 &  &  Galaxies \\
\hline
Blue early-type galaxies & 204 & 100\% &\\
\hline
Starforming & 50 & 24.5\%\\
AGN-SF composite & 52 & 25.5\%\\
Seyfert & 15 & 7.4\%\\
LINER & 11 & 5.4\%\\
Weak emission lines$^1$ & 76 & 37.3\%\\
\hline
\end{tabular}
\\$^1$ Measurement of at least one of H$\alpha$, H$\beta$, [NII] and
[OIII] has signal-to-nosie less than 3; see Section \ref{sec:eml}. \\
\end{center}
\end{table}

\subsection{Emission Line Measurement}
\label{sec:eml}
\begin{figure}
\begin{center}

\includegraphics[angle=90, width=0.49\textwidth]{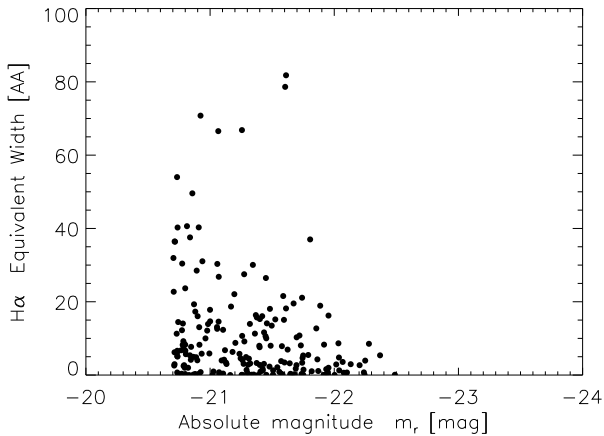}
\caption{The H$\alpha$ equivalent width as a function of galaxy absolute magnitude to give an indication of
the emission line strength.  \label{fig:ewa}}

\end{center}
\end{figure}

Optical emission lines are a well-established and powerful tool for
identifying and separating starforming galaxies from AGN
\citep{1981PASP...93....5B,1987ApJS...63..295V}. We measure the
optical emission lines of our blue early-type galaxy sample using the
\texttt{GANDALF} package\footnote{Available at :
\texttt{http://www.strw.leidenuniv.nl/sauron/}}
\citep{2006MNRAS.366.1151S}. In Figure \ref{fig:example_spec} we show
four example spectra of blue early-type galaxies processed with
\texttt{GANDALF}. We use the main emission lines ([OIII]
$\lambda$5007, H$\beta$, H$\alpha$, [NII] $\lambda$6583, [SII]
$\lambda$6583, [OI] $\lambda$6583) measured from such processed
spectra on diagnostic diagrams
\citep{2003MNRAS.346.1055K,2003ApJ...597..142M, 2001ApJ...556..121K,
2006MNRAS.372..961K}.  We classify all galaxies where the measurements
of H$\alpha$, H$\beta$, [NII] and [OIII] have a signal-to-noise of at
least 3 and divide them into purely starforming, SF-AGN composites and
AGN (Seyfert or LINER by comparing their line ratios on a BPT
diagram). Where available, we use the [OI] diagram for the
Seyfert-LINER separation.  If [OI] is unavailable, then we use [SII]
and if that also is not available, the [NII] diagram. To give an idea of the
emission line strength, we plot the H$\alpha$ equivalent width as a function 
of galaxy luminosity in Figure \ref{fig:ewa}.

\begin{figure}
\begin{center}

\includegraphics[angle=90, width=0.5\textwidth]{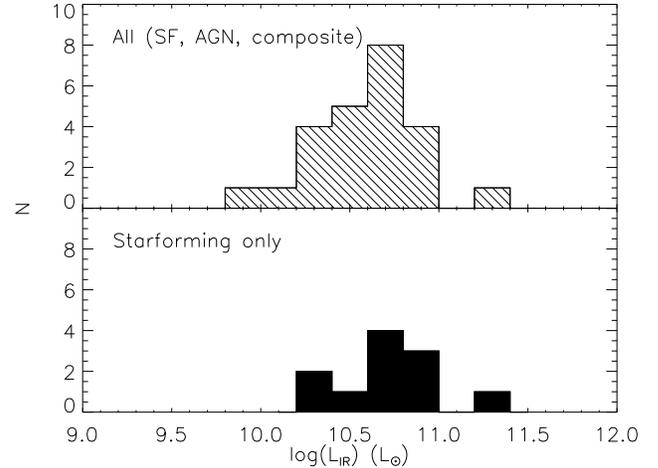}
\caption{The distribution of infrared luminosities $\rm L_{\rm IR}$
  for our blue early-type sample. The top panel is for the entire blue
  early-type galaxy sample, while the bottom shows the starforming
  objects only. The $\rm L_{\rm IR}$ of all objects may in part be due
  an AGN contribution, while the purely starforming sample is likely
  infrared luminous mostly due to star formation.}

\label{fig:l_ir}

\end{center}
\end{figure}

\begin{figure*}
\begin{center}

\includegraphics[angle=90, width=0.9\textwidth]{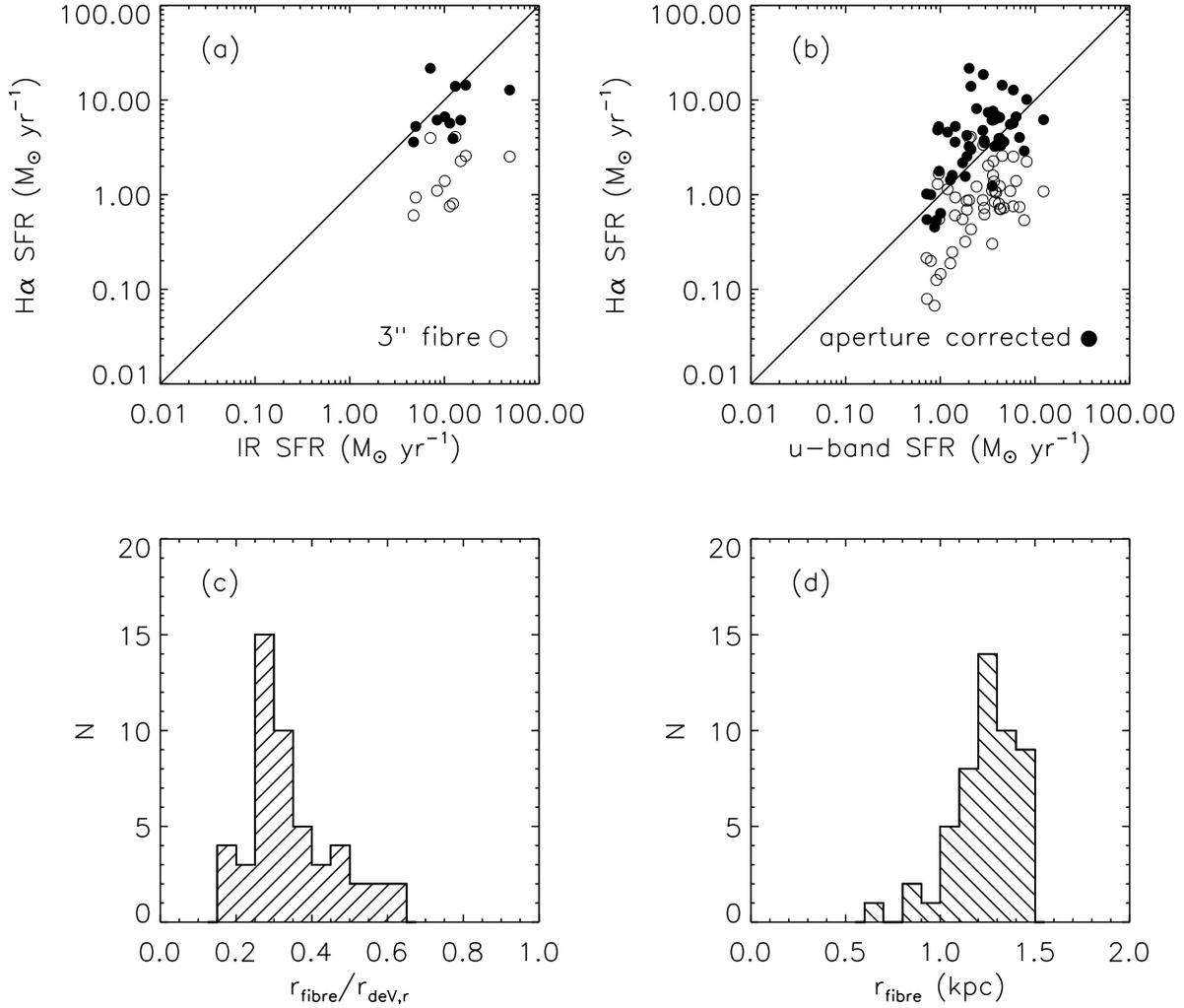}
\caption{In panels (a) and (b), we compare our various SFR indicators
  against each other for those blue early-type galaxies classified as
  purely starforming. In each, the empty circles stand for the
  3\arcsec fibre aperture H$\alpha$ SFR, while the filled circles
  represent the aperture-corrected H$\alpha$ SFR. In (a), we compare
  these two to the IR SFR and in (b) to the $u$-band SFR. We also
  indicate the line of the 1:1 correspondence. In the case of panel
  (a), we find that the 3\arcsec fibre SFR is systematically offset
  from the \textit{total} IR SFR, while the aperture-corrected SFR
  gives a good match. In the case of panel (b), the 3\arcsec fibre SFR
  again is systematically lower than the $u$-band SFR, while the
  aperture-corrected SFR is systematically higher. From this
  comparison, we conclude that while neither 3\arcsec fibre nor
  aperture-corrected SFR is perfect, the aperture-corrected SFR is
  closer to the IR SFR. In panel (c), we show the histogram of the
  ratio of the 3\arcsec fibre to the de Vaucouleurs scale radius and
  in (d), we show the histogram of physical radii in kpc that the
  3\arcsec fibre corresponds to.}

\label{fig:sfr_comparison}

\end{center}
\end{figure*}

In Figure \ref{fig:bpt_plots} and Table \ref{tab:class_results}, we
show our results. Only 25\% of the blue early-type galaxies are
unambiguously classified as actively starforming (corresponding to
1.5\% of the entire early-type galaxy population).

Interestingly, we find that 37\% of blue early-type galaxies do not
have $\rm S/N > 3$ in at least one of the four lines. We call this
class `weak emission line' galaxies. The lines most likely to drop
below $\rm S/N = 3$ are H$\beta$ and [OIII]. Out of the 76 galaxies
classified as quiescent, 32 have three lines out of four lines. This
lack of emission lines means that those galaxies may be no longer
strongly starforming or may not be host to a powerful AGN and that the
source of ionising radiation driving the emission lines is weak.  

It is possible to determine an upper limit on the current SFR of our
unclassifiable blue early-type galaxies by using whatever H$\alpha$
line flux is measured by \texttt{GANDALF}. We find that they have SFRs
$< 1~ \rm M_{\odot}yr^{-1}$; this represents the low-SFR tail of the
sample (see Section \ref{sec:sfr}). S07 found that blue early-type
galaxies typically undergo a burst in which 1-10\% of their stellar
mass are formed on a timescale of $\sim 100$ Myr. These blue
early-type galaxies without emission lines may represent the passive
evolutionary phase which follows this burst of star formation. They
may represent the population of early-types where the suppression of
star formation is complete and they are rapidly moving towards the red
sequence (S07). They may be distinct from normal passive early-type
galaxies in their UV/optical colours as demonstrated by
\cite{2005ApJ...619L.111Y}. \cite{2006ApJ...649L..71K,
2007ApJ...669..776K} discuss a potentially similar sample of
early-type galaxies with suppressed star formation at high redshift.

\begin{figure*}
\begin{center}

\includegraphics[angle=90, width=0.49\textwidth]{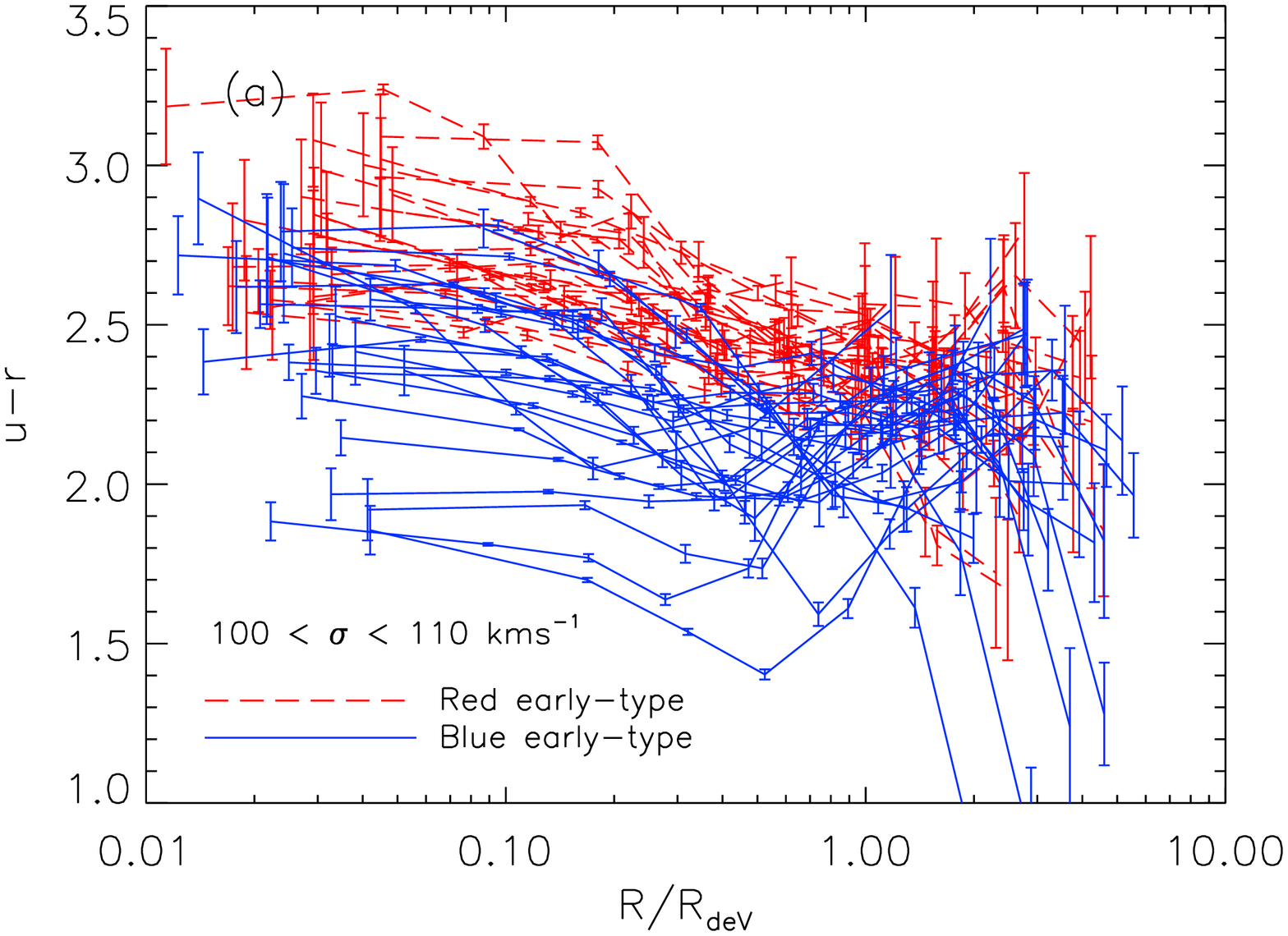}
\includegraphics[angle=90, width=0.49\textwidth]{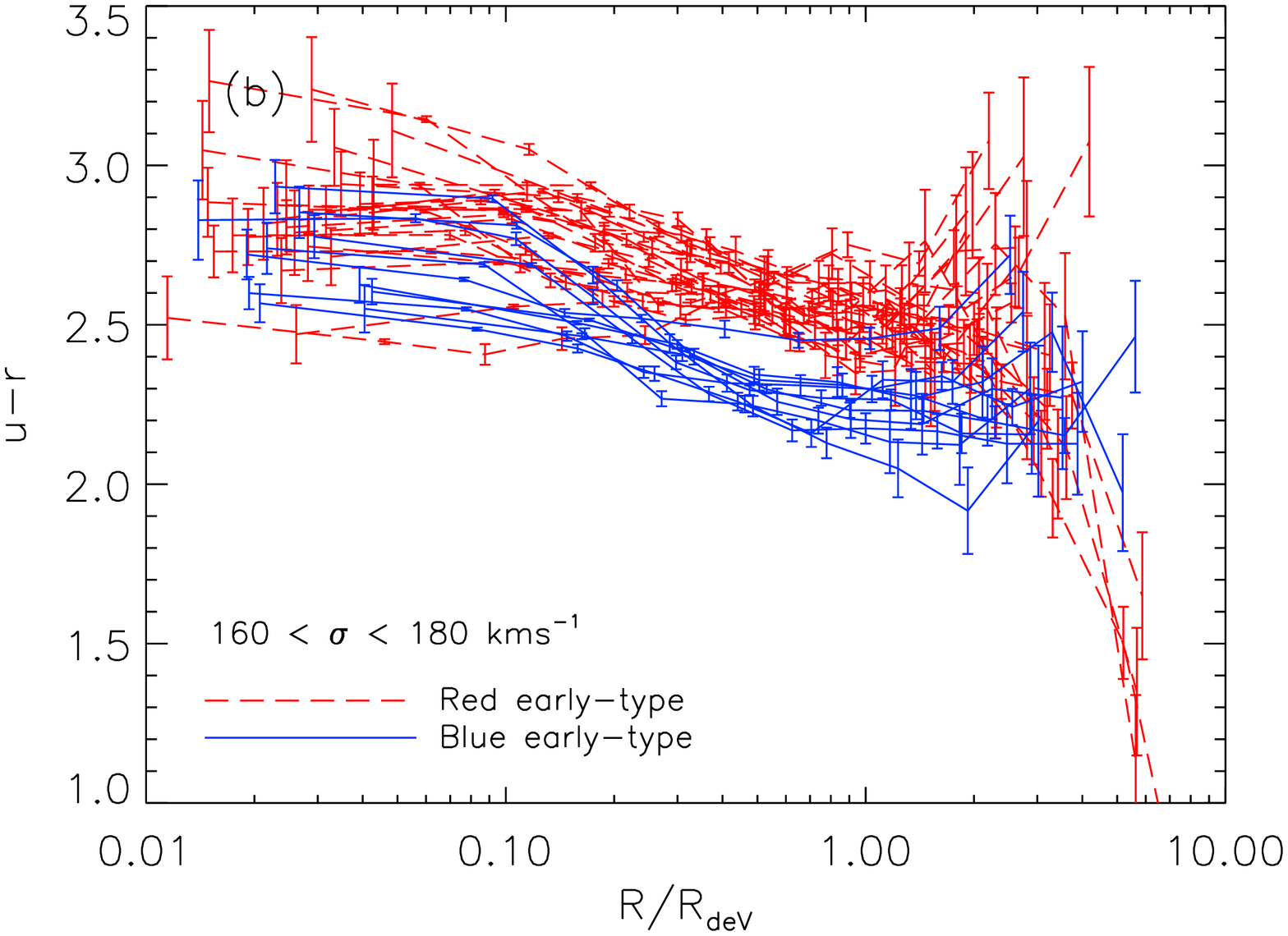}
\includegraphics[angle=90, width=0.49\textwidth]{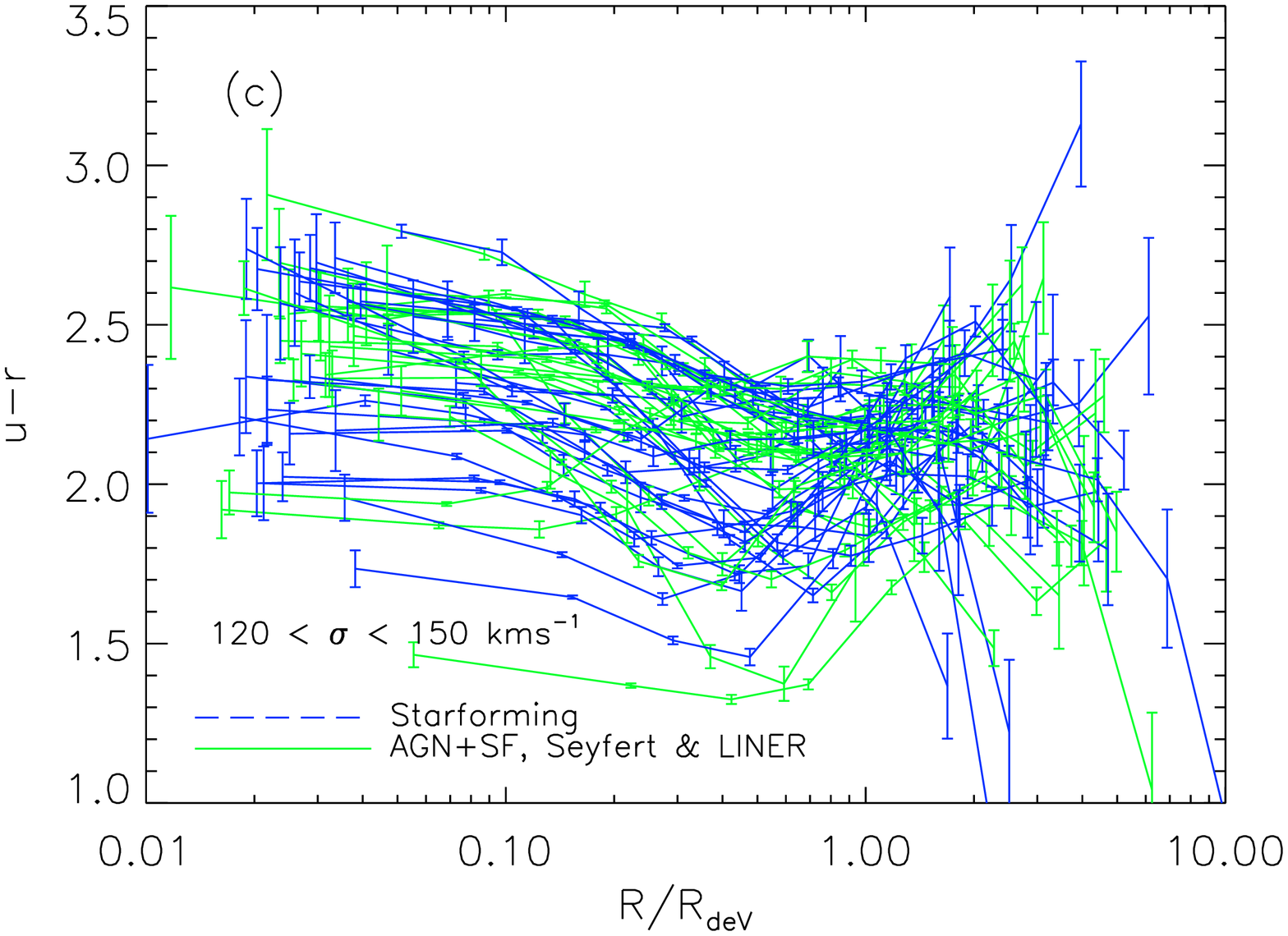}
\caption{The $u-r$ colour profiles of blue and red early-type galaxies. The radial distance is normalised
to the de Vaucouleurs scale radius derived by the SDSS photometric pipeline. In panels a and b, we show the colour profiles of blue and red early-types in two velocity dispersion bins. In both cases, we find
that the blue colours extend to at least the scale radius, supporting the idea that the star formation in
this population is spatially extended. In panel c) we test whether there is any clear systematic difference
in the colour profile between the purely starforming early-types (in blue), and those showing evidence for an AGN in the emission lines (AGN+SF, Seyfert and LINER; in green).}

\label{fig:radial}

\end{center}
\end{figure*}

\begin{figure}
\begin{center}

\includegraphics[angle=90, width=0.5\textwidth]{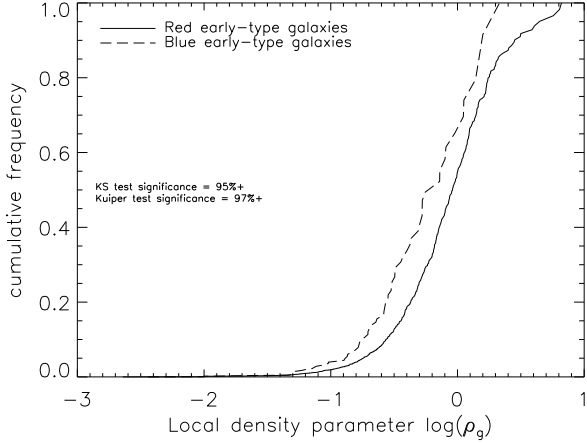}
\caption{We show cumulative frequency distribution of the local
  density parameter $\rmn{log(}\rho_g)$ (see
  \citealt{2007ApJS..173..512S}) for both blue early-type galaxies and
  their red counterparts in the same velocity dispersion regime ($110
  < \sigma < 150 ~ \rm kms^{-1}$ in order to avoid selection bias
  with mass. We use both a KS and a Kuiper test to assess the
  signifiance of the different distributions and find that they are
  different at 95\% and 97\% siginficance, respectively. Blue
  early-type galaxies prefer lower density environments comapred to
  their red counterparts.}

\label{fig:environment}

\end{center}
\end{figure}

\begin{figure}
\begin{center}

\includegraphics[angle=90, width=0.5\textwidth]{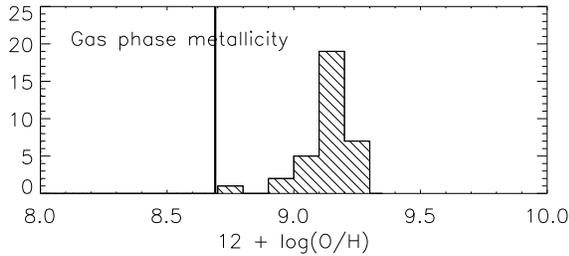}
\caption{This Figure shows the gas phase
  metallicity of the starforming early-types from the catalog of
  \citet{2004ApJ...613..898T}. The vertical line
  indicates the solar value \citep{2001ApJ...556L..63A}.}

\label{fig:sfr_met}

\end{center}
\end{figure}

\begin{figure}
\begin{center}

\includegraphics[angle=90, width=0.5\textwidth]{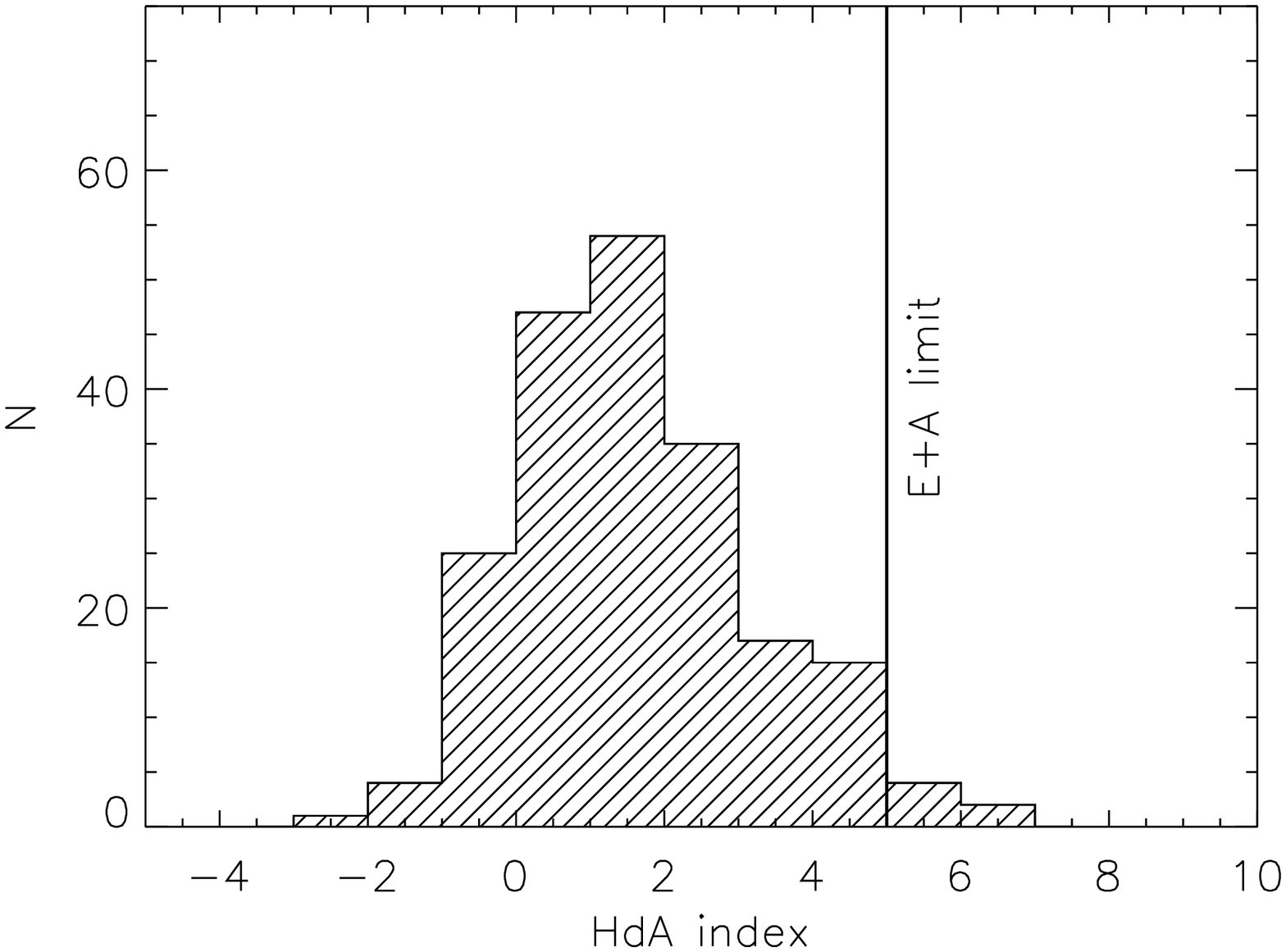}
\caption{Histogram of the HdA line index for the blue early-type
  galaxy sample. We indicate the E+A galaxy limit of 5\AA.}

\label{fig:hda_index}

\end{center}
\end{figure}

\section{The Properties of Blue Early-type Galaxies}
We now discuss the properties of the blue early-type galaxy population
in detail. We note that in our discussion of star formation, we
\textit{only} refer to those blue early-type galaxies classified as
purely starforming on the emission line diagrams (see Section
\ref{sec:eml}).

\subsection{Infrared Luminosities and Star Formation Rates}
\label{sec:sfr}

We match our blue early-type galaxy sample to the \textit{IRAS}
(\textit{Infrared Astronomical Satellite}) Faint Source Catalog v.2
\citep{1990BAAS...22Q1325M} to measure their infrared luminosities
$\rm L_{\rm IR}$ \citep{1996ARA&A..34..749S}, yielding 24 matches.  We
find that all but two of those early-type galaxies that are detected
range between $10^{10} < \rm L_{\rm IR} < 10^{11} L_{\odot}$ and two
objects are above the limit of luminous infrared galaxies (LIRGs; $\rm
L_{\rm IR} > 10^{11} \rm L_{\odot}$). In Figure \ref{fig:l_ir}, we
plot the distribution of infrared luminosities split into purely
starforming blue early-types and everything else.  We calculate IR
(infra-red) SFRs for those galaxies in our sample that have been
detected by IRAS using the calibration of \cite{2004ApJ...606..271G}
and find that the IR SFRs range between $5 < \rm SFR  < 50~ \rm
M_{\odot}yr^{-1}$. Using the relationships of
\cite{2004ApJ...606..271G}, we can also estimate the range of HCN
luminosity ($10^{7} < L_{\rm HCN} < 10^{8}~ \rm K~ kms^{-1}~pc^{2}$)
and CO luminosity ($10^{8.5} < L_{\rm CO} < \rm 10^{9.5}
K~kms^{-1}~pc^{2}$).  The IRAS FSC has a very shallow flux limit, so
that any detections are due to the most luminous and nearest
objects. In order to study the population as a whole, we turn to
H$\alpha$ as a tracer of the SFR.

\subsection{Star Formation Rates}
We measure the H$\alpha$ line luminosity, correcting for internal
extinction based on the Balmer decrement, and compute the star
formation rate using the calibration of \cite{1998ApJ...498..541K} for
our purely starforming blue early-type sample. We ignore the AGN+SF
composites, as we cannot know what fraction of the H$\alpha$ derives
from star formation. Since SDSS spectroscopic fibres have a 3\arcsec
diameter, this yields the SFR in the central 3\arcsec of the galaxy,
which corresponds for this sample to a mean of 2 kpc in radius.
Assuming that the distribution of the star formation follows the
stellar light of the galaxy, we can estimate the H$\alpha$ luminosity
lost outside the fibre from the ratio of the $r$-band flux within the
3\arcsec and the total Petrosian $r$-band flux. This typically scales
the light from $\sim 20\%$ inside the fibre to 100\%. In order to
determine whether this aperture correction is appropriate, we compare
with independent tracers of the total SFR: the IR SFR from the
\textit{IRAS} observations and the SFR from the $u$-band light
\citep{2003ApJ...599..971H}, corrected for extinction using the Balmer
decrement. Our galaxies are effectively point sources for IRAS and so
include emission from the entire galaxy.

In Figure \ref{fig:sfr_comparison}, we show the comparison between our
SFR indicators. While there is significant scatter, the aperture
corrected SFR is generally in better agreement with the IR SFR than
the fibre SFR and the $u$-band SFR lies between the aperture-corrected
and fibre SFR. This means that the central fibre SFR only samples a
fraction of the total SFR, which implies that star formation in blue
early-type galaxies is not confined to a central disk or nuclear
starburst, but extends over the entire galaxy (see also
\citealt{2006MNRAS.366.1151S}). 

Even with the correction, there is still an offset between the
H$\alpha$ and $u$-band SFRs. This may be due to the fact that the
$u$-band does not probe sufficiently obscured SF, while the IR
does. This underestimation of the SFR by the $u$-band light is thus
unsurprising. The better agreement between aperture-corrected
H$\alpha$ and IR SFR argues that H$\alpha$ suffers less from this
effect.

For the remainder, we use the aperture-corrected H$\alpha$ SFR which
is available for the entire starforming sample, unlike the IR SFR.

\subsection{Colour Profiles}
Another way to probe the distribution of young stars in our galaxies is to plot
the $u-r$ colour as a function of radius. In Figure \ref{fig:radial}, we plot the $u-r$ 
radial profiles for blue and red early-types in two velocity dispersion bins. Comparing at
the same velocity dispersion is important, but at low velocity dispersions, it can be challenging (c.f. Section \ref{sec:fj}). At low velocity dispersions, most passive, red early-types are not included in our 
volume limited sample with a bias against the redder, less luminous objects. We nevertheless plot
the colour profiles for a sample with $100 < \sigma < 110 ~ \rm kms^{-1}$ (Fig. \ref{fig:radial}a) and for $160 < \sigma < 180 ~ \rm kms^{-1}$ (Fig. \ref{fig:radial}b). In both cases, we find that the $u-r$ colour is significantly bluer out to at least the effective radius, supporting our finding that the star formation is extended, rather than circumnuclear, as has been found for some massive radio galaxies \citep{2001MNRAS.325..636A,2007MNRAS.381..611H}. Some blue early-types do become redder
at about $R_{\rm deV}$, others remain blue out to several radii. Some of the massive blue early-types
in Figure \ref{fig:radial}b tend to red $u-r$ colour within the central $1/10 ~ R_{\rm deV}$; this may be due to high levels of extinction.

We also test whether there is any clear difference between the purely starforming blue early-type 
galaxies and those with any evidence for AGN activity (i.e. AGN+SF, Seyfert and LINER). We
plot the colour profiles in Figure \ref{fig:radial}c and conclude that there is no clear trend apparent.

\subsection{Gas Phase Metallicities}
The chemical enrichment of the interstellar medium (ISM) of galaxies
is an important probe of their chemical evolution. In order to place
blue early-type galaxies in context, we obtain gas phase metallicities
from the SDSS catalogue of \cite{2004ApJ...613..898T} for those blue
early-type galaxies classified as purely starforming. We show the
distribution of gas-phase metallicities for those objects with matches
in their catalogue in Figure \ref{fig:sfr_met}
\citep{2004ApJ...613..898T}. We find that the gas phase metallicities
are all supersolar (adopting the solar value of
\citealt{2001ApJ...556L..63A}).

From the BPT diagrams in Figure \ref{fig:bpt_plots}, we can see that there is a lack of
objects with high [OIII]/H$\beta$ ratios on the starforming locus is naturally accounted
for by the lack of low metallicity gas in blue early-types. Comparing to the observed
mass-metallicity relation of \cite{2004ApJ...613..898T}, the gas phase metallicities
of blue early-type galaxies are as expected and the high metallicities are a result of our
luminosity cut.

\subsection{Environment}
The effect of environment on galaxy formation has seen vigorous debate
since the discovery of \cite{1974ApJ...194....1O} and
\cite{1980ApJ...236..351D} that early-type galaxies preferentially
reside in clusters. It is not clear whether in particular the scaling
relations of the stellar populations of early-type galaxies vary at
all with environment.  We measure the local density of our sample
using the environment parameter $\rho_{g}$ defined by
\cite{2007ApJS..173..512S} (see also
\citealt{2008ApJS..176..414Y}). This parameter samples the local
galaxy distribution with a Gaussian kernel of 2 Mpc and represents a
weighted local number density. We calculate $\rho_{g}$ for our blue
early-type galaxy sample and for the corresponding red early-type
galaxies and show the resulting cumulative distributions in Figure
\ref{fig:environment}. In order to avoid bias with galaxy mass, we
limit both the blue and red early-type sample to a range of $110 <
\sigma < 150~ \rm kms^{-1}$.

The results for the two distributions of $\rho_{g}$ show that blue
early-type galaxies clearly reside in lower-density environments
compared to their red counterparts at the same velocity dispersion. We
perform a KS test and find a probability of the two distributions
being drawn from different parent distributions of $95\%$. For a
Kuiper test, the signifiance is $97\%$. In fact, blue early-type
galaxies appear to be virtually absent at the typical densities of
cluster centres as defined by a commonly used cluster catalogue based
on SDSS, the C4 cluster catalogue of \cite{2005AJ....130..968M} which
were explored by \cite{2007ApJS..173..512S}.

\section{Discussion}
Blue early-type galaxies represent a fascinating phase in the
evolution of galaxies. Quite what place they take in the overarching
picture is still unclear. There are three general scenarios that might
account for them. They could represent the result of a spiral-spiral
merger whose end-product will eventually join the red sequence in
passive evolution. They might also be early-type galaxies on the red
sequence that are undergoing an episode of star formation due to the
sudden availability of cold gas, making them leave the red sequence
before rejoining it. The third possibility is that they are mixed
early-type/late-type mergers \citep[e.g.][]{2003ApJ...597L.117K},
meaning that one progenitor originates from each the blue cloud and
the red sequence. Which of these scenarios -- or which combination
thereof -- is ocurring here is unclear. We provide Table \ref{tab:cat}
to enable follow-up observations of these objects to further
illuminate their nature.

\subsection{Are there any Nearby Counterparts?}
In the nearby Universe, our SDSS blue star-forming early-type galaxies
could correspond to objects like NGC~3032, which is the bluest of the
early-type galaxies observed in the course of the SAURON
integral-field survey \citep{2002MNRAS.329..513D} and the one showing
by far the strongest nebular emission \citep{2006MNRAS.366.1151S}.
NGC~3032 is also unique in that it shows star formation across the
entire SAURON field-of-view (corresponding to approximately one
effective radius), whereas in early-type galaxies with lower levels of
star formation, this occurs in circumnuclear rings. The equivalent
width of the integrated nebular emission of NGC~3032 ($4\rm \AA$
within one effective radius for H$\beta$) is also sufficiently strong
that it would have been detected by SDSS observations and end up as
being classified as one of our star-forming blue early-type galaxies.

Arguing from population statistics, that only one such a galaxy is
found among the 48 objects surveyed by SAURON is also consistent with
the 1.5\% fraction of the early-type population that we found to be
blue and star-forming.

\subsection{Comparison to Theoretical Models}
We now explore the blue early-type galaxy population in simulations of
galaxy formation.

\subsubsection{Semi-Analytic Models}
In order to compare our results with the predictions of the
$\Lambda$CDM paradigm we use semi-analytical modelling (SAM)
\citep{2005MNRAS.359.1379K,2006MNRAS.370..902K} including an empirical
description for AGN-feedback \citep{2006Natur.442..888S}.  This model
reproduces well the fraction of NUV-blue early-type galaxies as a
function of magnitude and velocity dispersion
\citep{2006Natur.442..888S}. It should be noted that without our
AGN-feeedback prescription almost all of the simulated early-type
galaxies were blue in the luminosity range of interest; reproducing
the colours of early-type galaxies remains a difficult challenge for
SAMs. We here compare the range of star formation rates for the blue
early-type population with those obtained from the observations.

Adopting a magnitude limit of $M_r=-21$ and the same $3\sigma$ cut in
$u-r$ as described above to select blue early-type galaxies, we find
that 7.7\% of model early-type galaxies are selected as blue. This
fraction is comparable to the $5.7\pm0.4\%$ that we observe (see
Figure \ref{fig:optical_cmr}); the difference may be due to
massive gas-rich mergers that are considered bulge-dominated objects
in the model, but may have sufficiently disturbed morphologies to have
been classified as mergers in Galaxy Zoo. In the blue early-type
galaxies in the model, we find SFRs ranging from $0.1-45 \rm
~M_{\odot}yr^{-1}$, with a strong peak around $1 \rm
~M_{\odot}yr^{-1}$ and an extended tail to the upper end of the SFR
range. The fraction of blue early-type galaxies showing SFR above $1
\rm ~M_{\odot}yr^{-1}$ is $\sim 45 \%$ and as such is larger than the
above reported 25\%. However, the combined starforming and AGN-SF
composite population accounts for $\sim$ 50\% of the galaxies and is
well above the fraction of starforming early-types in the SAM,
indicating that a large fraction of the AGN-SF composite population
might indeed be starforming.

\subsubsection{A Merger Trigger for Blue Early-types?}
\cite{2007arXiv0711.1493K} have performed numerical simulations of
minor mergers to determine their role in the recent star formation
histories of early-type galaxies and explore whether satellite
accretion could produce the blue UV colours that seem to be common
amongst the low redshift early-type population
\citep{2005ApJ...619L.111Y, 2007ApJS..173..512S,
2007ApJS..173..619K}. The peak SFRs produced in minor mergers (with
mass ratios of 1:6 and a satellite with a gas fraction of 20\%) is
less than $\sim 1~ \rm M_{\odot}yr^{-1}$ (see their Figure 1). Even if
larger mass ratios (e.g. 1:4) and more gas-rich satellites (e.g. with
gas fractions of 40\%) are considered the peak SFRs in these events
are unlikely to be higher than a few $\rm M_{\odot
}yr^{-1}$. Therefore, it is unlikely that the star formation activity
observed in our sample is triggered by minor mergers.

Accretion of cold gas from the halo should be negligible if feedback
processes efficiently maintain the temperature of the hot gas
reservoir in early-type galaxies \citep{2004MNRAS.347.1093B}, but it
cannot be ruled out. A plausible route for the triggering of the star
formation we are witnessing in these blue early-type systems is
gas-rich major mergers, where at least one of the progenitors is a
late-type galaxy. Such events are predicted by current semi-analytic
models \citep{2003ApJ...597L.117K, 2006astro.ph..2347K} and so may be
a natural trigger for the the time sequence of S07.

\subsection{Are Blue Early-type Galaxies Progenitors of E+A Galaxies?}
\label{sec:ea}
An intriguing possibility worth discussing here is that the blue
early-type galaxy population may be at least partially linked to the
E+A galaxy phenomenon via an evolutionary sequence. E+A (or k+a)
galaxies have spectra undergoing no ongoing star formation indicated
by a lack of H$\alpha$ and [OII] emission lines but indicating a
recent past starburst resulting in a considerable mass-fraction in
intermediate age stars (the `A' in E+A denotes A stars). First noted
by \cite{1983ApJ...270....7D}, E+A galaxies are rare and seem to
reside mostly in low-density environment and their deep Balmer lines
together with their UV-optical colours point to a pathway in which
star formation was truncated on short timescales
\citep{2007MNRAS.382..960K}.

The usual selection of E+A galaxies is via a rejection of spectra with
signs of active star formation and extremely deep Balmer lines. If we
ignore the star formation criteria and simply focus on the Balmer line
cut, only a small fraction of blue early-type galaxies (6/204, 2.9\%,
see Figure \ref{fig:hda_index}) would meet the $\rm HdA > 5 \AA$
criterion. However, blue early-types are still actively starforming
resulting in smaller typical HdA equivalent widths (EW); the HdA EW
peaks at intermediate ages. Thus, it is conceivable that if star
formation is suddenly suppressed (e.g. via AGN feedback, S07 and \citealt{2009ApJ...690.1672S, 2009arXiv0901.1663S}), then a
part of the blue early-type galaxy population may represent the
progenitors of E+A galaxies. The deep Balmer lines apparent in the spectra
of blue early-types (see Figure \ref{fig:example_spec}) indicate that a
substantial intermediate-age population may be present.

\section{Conclusions}
In this Paper, we investigate a volume-limited sample of visually
identified early-type galaxies selected from Galaxy Zoo and SDSS
DR6. Our selection via morphology only has allowed us to study the
population of \textit{optically blue} early-type galaxies that do not
reside on the red sequence like their passive counterparts. We
determine the number fraction of these blue early-types and discuss
selection effects. The most prominent caveat is the fact that absolute
magnitudes can be boosted significantly, giving the impression that
blue early-type galaxies are massive. However, their velocity
dispersions are moderate and extremely few exceed $200 \rm ~kms^{-1}$,
implying  masses below $10^{11} M_{\odot}$. There are almost no
\textbf{massive} blue early-type  galaxies in the low redshift
Universe. We show that the young stellar populations present in blue
early-type galaxies cause them to scatter off the Faber-Jackson
relation and using simple fading/ageing models determine that they can
fade back onto the Faber-Jackson relation within $\sim 1$ Gyr for
small young mass fractions of a few percent.

We use emission line diagnostic diagrams to classify them into purely
starforming (25\%), SF+AGN composites (25\%) and AGN (12\%). The
remaining blue early-type galaxies show no strong emission lines and
so have suppressed their star formation and show no AGN (as discussed
in S07). We discuss a possible connection between these objects and
E+A/K+A galaxies in Section \ref{sec:ea}.

Using a variety of measures (H$\alpha$, $u$-band and IR), we determine
the range of star formation rates in blue early-type galaxies and find
a wide range from $0.5 < \rm SFR < 50~ M_{\odot}yr^{-1}$. The higher
values are the highest values reported for low redshift early-type
galaxies. By comparing the star formation rate probed by the central
SDSS 3$\arcsec$ fibres to total IR star formation rates, we conclude
that the star formation in starforming blue early-type galaxies is not
confined to a central circumnuclear starburst, but extends spatially
across most of the galaxy. We independently confirm this by verifying that
the radial colour profiles exhibit blue colours out to at least the de Vaucouleurs
scale radius, and in some cases, beyond.
We also measure the gas phase metallicity
and find them to be supersolar in every case, consistent with the
mass-metallicity relation of \cite{2004ApJ...613..898T}.

We measure the local environment using the measure of
\cite{2007ApJS..173..512S} and \cite{2008ApJS..176..414Y} and find, at
a given velocity dispersion, a statistically significant difference
between red and blue early-type galaxies. Blue early-type galaxies are
less frequent at the highest densities compared to their red
counterparts at the same mass, implying a role of environment. 

We discuss the place of these objects in our current understanding of
galaxy formation. A comparison to numerical simulation of minor
mergers suggests that minor mergers are an unlikely trigger mechanism;
this leaves major mergers (either early-early or early-late) and major
cooling events in red early-type galaxies as plausible
triggers. Further observations of blue early-type galaxies should be
able to resolve this. Deep optical imaging should reveal tidal tails
and other features if they are the result of major mergers (c.f
\citealt{2005AJ....130.2647V}).

In order to allow the further study of this exciting population, we
provide a catalog of all blue early-type galaxies presented in this
paper in Table \ref{tab:cat}.

\section{Acknowledgements}
We would like to thank Adrienne Slyz and Julien Devrient for helpful
comments and suggestions. We also would like to thank Alice Sheppard
and Edd Edmondson for their help in administering the Galaxy Zoo
forum. KS is supported by the Henry Skynner Junior Research Fellowship
at Balliol College Oxford. CJL acknowledges support from the STFC
Science in Society Programme. S. Kaviraj acknowledges a Leverhulme
Early-Career Fellowship, a BIPAC fellowship and a Research Fellowship
from Worcester College, Oxford. This work was supported by grant
No.~R01-2006-000-10716-0 from the Basic Research Program of the Korea
Science and Engineering Foundation to SKY.

Funding for the SDSS and SDSS-II has been provided by the Alfred
P. Sloan Foundation, the Participating Institutions, the National
Science Foundation, the U.S. Department of Energy, the National
Aeronautics and Space Administration, the Japanese Monbukagakusho, the
Max Planck Society, and the Higher Education Funding Council for
England. The SDSS Web Site is http://www.sdss.org/.

The SDSS is managed by the Astrophysical Research Consortium for the
Participating Institutions. The Participating Institutions are the
American Museum of Natural History, Astrophysical Institute Potsdam,
University of Basel, University of Cambridge, Case Western Reserve
University, University of Chicago, Drexel University, Fermilab, the
Institute for Advanced Study, the Japan Participation Group, Johns
Hopkins University, the Joint Institute for Nuclear Astrophysics, the
Kavli Institute for Particle Astrophysics and Cosmology, the Korean
Scientist Group, the Chinese Academy of Sciences (LAMOST), Los Alamos
National Laboratory, the Max-Planck-Institute for Astronomy (MPIA),
the Max-Planck-Institute for Astrophysics (MPA), New Mexico State
University, Ohio State University, University of Pittsburgh,
University of Portsmouth, Princeton University, the United States
Naval Observatory, and the University of Washington.

\bibliographystyle{mn}

\bsp

\begin{deluxetable}{lcclllccc}
\tablecolumns{8}
\tablewidth{0pc}
\tablecaption{A catalogue of blue early-type galaxies}
\tablehead{
 \colhead{SDSS Object ID} & 
 \colhead{RA} &
 \colhead{DEC} &  
 \colhead{Redshift} &  
 \colhead{$M_{r}$} &
 \colhead{$u-r$} & 
 \colhead{Star Formation Rate$^1$} & 
 \colhead{Emission Line Class$^2$} \\
 \colhead{} &
 \colhead{} &
 \colhead{} &
 \colhead{} &
 \colhead{} &
 \colhead{} &
 \colhead{$(M_{\odot}yr^{-1})$} &
 \colhead{} 
}
\startdata
      587722981736906813 &   11:30:57.9 &-01:08:51.1 &      0.04802 &     -21.43 &      2.365 &          - &    Seyfert\\
       587722982272925748 &   11:23:27.0 &-00:42:48.8 &      0.04084 &     -20.81 &      1.606 &        4.5 &         SF\\
       587722982276923512 &   12:00:04.3 &-00:45:34.5 &      0.04711 &     -21.08 &      2.088 &          - &     AGN+SF\\
       587722982299271230 &   15:23:47.1 &-00:38:23.0 &      0.03747 &     -21.00 &      1.998 &        2.5 &         SF\\
       587722983351582785 &   12:08:23.5 &+00:06:37.0 &      0.04081 &     -21.50 &      2.038 &        3.5 &         SF\\
       587722984440463382 &   14:26:13.0 &+00:51:38.1 &      0.03193 &     -20.87 &      2.035 &          - &      LINER\\
       587724198282133547 &   01:41:43.2 &+13:40:32.8 &      0.04539 &     -21.81 &      1.432 &        12. &         SF\\
       587724199351812182 &   01:03:58.7 &+15:14:50.1 &      0.04176 &     -21.40 &      2.261 &        7.0 &         SF\\
       587724199355285539 &   01:36:19.4 &+14:39:24.6 &      0.03369 &     -20.80 &      2.310 &          - &     AGN+SF\\
       587724231570817120 &   01:44:26.2 &+13:09:35.5 &      0.04502 &     -20.93 &      2.152 &          - &     AGN+SF\\
       587724242298077193 &   02:25:22.1 &-07:51:06.1 &      0.03919 &     -20.79 &      2.243 &          - &     AGN+SF\\
       587724650327113796 &   11:26:40.7 &-01:41:37.6 &      0.04671 &     -21.95 &      1.857 &          - &      LINER\\
       587725491599835179 &   17:18:56.0 &+57:22:18.8 &      0.03160 &     -20.98 &      1.993 &          - &     AGN+SF\\
       587725550139408424 &   12:35:02.6 &+66:22:33.4 &      0.04684 &     -21.67 &      1.959 &        6.1 &         SF\\
       587725550675755071 &   12:22:49.1 &+66:51:45.3 &      0.03222 &     -21.66 &      2.270 &          - &          -\\
       587725551199322281 &   08:35:04.2 &+52:49:53.5 &      0.04454 &     -21.94 &      2.402 &          - &          -\\
       587725773459554601 &   08:07:01.7 &+45:42:29.3 &      0.04939 &     -21.56 &      2.361 &          - &          -\\
       587725773460733977 &   08:15:50.5 &+47:48:39.6 &      0.04029 &     -21.31 &      2.185 &          - &     AGN+SF\\
       587725979615559791 &   08:36:14.9 &+52:12:09.1 &      0.04809 &     -20.77 &      1.906 &          - &    Seyfert\\
       587725981226107027 &   08:29:09.1 &+52:49:06.9 &      0.04842 &     -21.00 &      1.835 &        2.1 &         SF\\
       587726014009573396 &   14:11:17.7 &+01:16:31.0 &      0.02499 &     -20.71 &      2.299 &          - &     AGN+SF\\
       587726016149454996 &   13:02:10.8 &+03:06:23.8 &      0.04724 &     -22.37 &      2.065 &          - &     AGN+SF\\
       587726031175548968 &   12:06:47.2 &+01:17:09.8 &      0.04124 &     -21.18 &      2.207 &        1.2 &         SF\\
       587726031691514020 &   08:55:50.3 &+00:49:52.1 &      0.04171 &     -20.90 &      2.031 &          - &     AGN+SF\\
       587726032249749558 &   12:11:10.8 &+02:15:19.9 &      0.01981 &     -20.71 &      2.266 &          - &          -\\
       587726032780066885 &   11:11:06.9 &+02:28:48.0 &      0.03514 &     -20.87 &      2.259 &          - &    Seyfert\\
       587726033307238543 &   09:42:07.7 &+02:28:05.1 &      0.04825 &     -21.40 &      2.361 &          - &    Seyfert\\
       587726033873862745 &   14:14:32.1 &+03:11:24.9 &      0.02691 &     -21.82 &      2.334 &          - &          -\\
       587726033878122590 &   14:53:40.0 &+02:55:14.4 &      0.04447 &     -20.84 &      1.876 &          - &     AGN+SF\\
       587726100953432183 &   15:17:19.7 &+03:19:18.9 &      0.03749 &     -20.74 &      2.131 &       0.73 &         SF\\
       587727178995204109 &   01:22:09.2 &-09:53:18.8 &      0.04273 &     -21.93 &      2.394 &          - &          -\\
       587727180061999117 &   00:17:34.9 &-09:32:36.1 &      0.02237 &     -21.58 &      2.388 &          - &          -\\
       587727180070715469 &   01:38:22.3 &-08:58:31.4 &      0.04240 &     -20.80 &      2.054 &          - &          -\\
       587727221414494275 &   00:09:07.9 &+14:27:55.8 &      0.04141 &     -21.42 &      2.219 &          - &    Seyfert\\
       587727221944746290 &   23:06:54.9 &+14:11:30.3 &      0.03992 &     -21.11 &      2.298 &          - &     AGN+SF\\
       587727221949005979 &   23:47:03.8 &+14:50:30.4 &      0.02217 &     -20.82 &      2.157 &          - &    Seyfert\\
       587727225693667361 &   00:32:45.2 &-10:27:03.6 &      0.03833 &     -21.51 &      2.353 &          - &          -\\
       587727227846983785 &   01:26:49.9 &-08:29:27.2 &      0.04968 &     -21.92 &      2.355 &          - &          -\\
       587727943497613319 &   10:04:38.7 &+01:53:22.5 &      0.03034 &     -21.02 &      2.324 &          - &          -\\
       587728664506794118 &   08:29:57.5 &+44:56:24.6 &      0.04271 &     -21.95 &      2.389 &          - &          -\\
       587728879266562060 &   10:16:28.4 &+03:35:02.7 &      0.04848 &     -21.72 &      2.380 &        6.1 &         SF\\
       587728905563603159 &   07:56:49.9 &+34:49:52.6 &      0.04096 &     -21.06 &      2.039 &          - &     AGN+SF\\
       587728920059183194 &   14:43:05.3 &+61:18:38.6 &      0.04796 &     -21.60 &      1.344 &          - &     AGN+SF\\
       587728947978436717 &   10:42:32.3 &-00:41:58.3 &      0.04981 &     -20.82 &      2.048 &          - &          -\\
       587728948511637597 &   10:08:47.8 &-00:16:11.3 &      0.04536 &     -22.07 &      2.239 &          - &     AGN+SF\\
       587728949049557028 &   10:18:06.7 &+00:05:59.7 &      0.04840 &     -21.65 &      2.307 &          - &    Seyfert\\
       587728949584658573 &   10:02:22.5 &+00:33:31.0 &      0.04424 &     -20.89 &      2.221 &          - &     AGN+SF\\
       587729157894832296 &   13:18:37.5 &+03:59:42.4 &      0.04574 &     -20.79 &      2.312 &          - &     AGN+SF\\
       587729159518945292 &   15:22:01.9 &+03:45:06.3 &      0.03714 &     -22.20 &      2.424 &          - &     AGN+SF\\
       587729160046510114 &   13:57:07.5 &+05:15:06.8 &      0.03967 &     -21.86 &      2.158 &        6.6 &         SF\\
       587729227152752660 &   16:03:44.5 &+52:24:12.5 &      0.04322 &     -20.74 &      2.216 &          - &          -\\
       587729227683659793 &   14:51:15.7 &+62:00:14.6 &      0.04306 &     -21.45 &      2.173 &        3.9 &         SF\\
       587729387675844667 &   08:51:35.6 &+47:33:27.6 &      0.02910 &     -22.25 &      2.415 &          - &     AGN+SF\\
       587729407542624367 &   16:43:23.1 &+39:48:23.2 &      0.03024 &     -21.75 &      2.350 &          - &    Seyfert\\
       587729408611647550 &   16:10:51.8 &+48:54:39.2 &      0.04504 &     -20.94 &      1.995 &          - &    Seyfert\\
       587729408622723373 &   17:23:24.9 &+27:48:46.3 &      0.04845 &     -21.46 &      2.017 &        3.2 &         SF\\
       587729751674323201 &   17:04:08.7 &+31:29:02.4 &      0.03304 &     -21.27 &      2.318 &          - &     AGN+SF\\
       587729778516820024 &   14:06:56.4 &-01:35:41.0 &      0.02916 &     -21.37 &      2.127 &        5.7 &         SF\\
       587729778520817884 &   14:43:18.6 &-01:15:12.5 &      0.04982 &     -21.63 &      2.364 &          - &          -\\
       587730773351006383 &   23:51:29.1 &+14:04:06.2 &      0.03938 &     -21.25 &      2.029 &          - &    Seyfert\\
       587730816286785593 &   22:15:16.2 &-09:15:47.6 &      0.03843 &     -21.61 &      1.707 &        21.0 &         SF\\
       587731172234428781 &   21:14:52.4 &-00:58:39.9 &      0.04402 &     -21.95 &      2.415 &          - &          -\\
       587731185117954089 &   22:08:06.1 &-00:54:25.0 &      0.03797 &     -20.77 &      2.167 &          - &          -\\
       587731186744950905 &   00:38:14.7 &+00:12:37.5 &      0.04440 &     -20.84 &      2.311 &          - &     AGN+SF\\
       587731187267403859 &   22:26:14.6 &+00:40:04.1 &      0.03642 &     -22.04 &      2.340 &          - &          -\\
       587731500261834946 &   10:54:37.9 &+55:39:46.0 &      0.04787 &     -20.89 &      1.976 &        6.5 &         SF\\
       587731501332430930 &   10:06:04.3 &+53:42:53.5 &      0.04427 &     -21.01 &      1.613 &          - &     AGN+SF\\
       587731512617402577 &   03:01:26.2 &-00:04:25.5 &      0.04285 &     -21.10 &      2.156 &        3.2 &         SF\\
       587731513696845940 &   03:53:34.8 &+00:48:50.5 &      0.03785 &     -20.79 &      1.753 &          - &     AGN+SF\\
       587731514229784683 &   03:17:49.3 &+01:13:37.2 &      0.03685 &     -21.26 &      2.200 &          - &    Seyfert\\
       587731514501365802 &   03:46:12.1 &+01:07:15.9 &      0.03805 &     -21.09 &      2.318 &          - &          -\\
       587731521207730241 &   09:13:23.7 &+43:58:34.2 &      0.04292 &     -21.45 &      2.176 &        14.0 &         SF\\
       587731680111099933 &   07:54:01.2 &+29:25:14.5 &      0.03608 &     -20.71 &      2.296 &          - &      LINER\\
       587731874461974650 &   08:12:25.5 &+37:43:48.8 &      0.03852 &     -21.07 &      1.448 &          - &     AGN+SF\\
       587731887347662946 &   08:21:49.6 &+39:14:49.6 &      0.02821 &     -20.74 &      2.022 &          - &     AGN+SF\\
       587731887888990283 &   09:00:36.1 &+46:41:11.4 &      0.02748 &     -21.30 &      2.200 &          - &          -\\
       587731889505632389 &   12:20:37.4 &+56:28:46.2 &      0.04381 &     -20.84 &      2.238 &       0.53 &         SF\\
       587732053246017714 &   09:18:55.2 &+42:00:13.0 &      0.04113 &     -21.74 &      2.425 &          - &          -\\
       587732054318645354 &   09:05:37.2 &+41:05:32.2 &      0.04137 &     -21.70 &      2.342 &          - &      LINER\\
       587732134844366940 &   11:06:21.4 &+50:23:16.8 &      0.03988 &     -20.91 &      2.011 &          - &    Seyfert\\
       587732135382089788 &   11:17:33.3 &+51:16:17.7 &      0.02767 &     -21.40 &      2.332 &          - &          -\\
       587732135921057870 &   11:47:41.7 &+52:26:55.8 &      0.04816 &     -21.92 &      2.226 &          - &     AGN+SF\\
       587732136457011245 &   11:33:39.1 &+52:40:28.6 &      0.04913 &     -20.80 &      2.163 &          - &          -\\
       587732157389013064 &   07:54:20.6 &+25:51:33.2 &      0.04167 &     -21.06 &      2.011 &        1.7 &         SF\\
       587732469849653424 &   08:29:30.0 &+31:40:31.7 &      0.01828 &     -20.73 &      2.090 &          - &     AGN+SF\\
       587732471463411876 &   08:53:11.4 &+37:08:06.5 &      0.04980 &     -21.59 &      2.063 &        10.0 &         SF\\
       587732484375838770 &   15:33:44.5 &+39:13:42.0 &      0.04034 &     -21.26 &      2.283 &          - &          -\\
       587732577238974514 &   11:07:16.6 &+06:18:08.9 &      0.02915 &     -22.24 &      2.502 &          - &          -\\
       587732580982521898 &   11:19:07.6 &+58:03:14.4 &      0.03250 &     -21.40 &      2.171 &          - &          -\\
       587732582592479319 &   11:05:59.0 &+58:56:45.8 &      0.04765 &     -22.02 &      2.435 &          - &    Seyfert\\
       587732701792895177 &   10:48:07.9 &+06:18:19.4 &      0.04196 &     -21.13 &      2.278 &          - &     AGN+SF\\
       587732769982906490 &   12:20:23.1 &+08:51:37.1 &      0.04894 &     -20.88 &      2.086 &        3.6 &         SF\\
       587732771580149928 &   10:16:58.0 &+08:39:16.1 &      0.02775 &     -20.92 &      2.334 &          - &          -\\
       587732772653957258 &   10:17:06.8 &+09:36:22.6 &      0.04393 &     -21.09 &      1.588 &          - &     AGN+SF\\
       587733081351258168 &   12:40:55.5 &+55:27:15.2 &      0.03175 &     -21.29 &      2.357 &          - &          -\\
       587733195160485975 &   13:05:25.8 &+53:35:30.3 &      0.03812 &     -21.68 &      2.212 &          - &          -\\
       587733197845561478 &   13:19:31.5 &+55:17:50.5 &      0.03813 &     -21.48 &      2.325 &          - &     AGN+SF\\
       587733397559902223 &   14:01:55.2 &+51:31:10.3 &      0.04122 &     -20.91 &      2.291 &          - &    Seyfert\\
       587733398642688031 &   15:47:44.1 &+41:24:08.3 &      0.03260 &     -21.04 &      2.156 &          - &    Seyfert\\
       587733410445131853 &   14:02:24.5 &+51:41:13.1 &      0.04120 &     -21.75 &      2.388 &          - &          -\\
       587733411518808182 &   14:02:48.8 &+52:30:00.8 &      0.04361 &     -20.77 &      1.893 &        1.4 &         SF\\
       587733412055941186 &   14:07:47.2 &+52:38:09.7 &      0.04381 &     -20.78 &      1.812 &        4.2 &         SF\\
       587733412064788620 &   15:50:00.5 &+41:58:11.2 &      0.03391 &     -20.80 &      1.879 &        3.6 &         SF\\
       587733412068261940 &   16:19:32.4 &+36:31:16.3 &      0.03636 &     -21.16 &      2.225 &          - &     AGN+SF\\
       587733432459788474 &   16:51:16.7 &+28:06:52.5 &      0.04724 &     -21.59 &      1.921 &        4.0 &         SF\\
       587733604808524062 &   16:56:20.3 &+32:10:27.4 &      0.03671 &     -21.32 &      2.340 &          - &     AGN+SF\\
       587734303266308145 &   22:05:15.4 &-01:07:33.4 &      0.03172 &     -21.38 &      2.011 &          - &      LINER\\
       587734622698602943 &   07:47:23.1 &+22:20:41.3 &      0.04549 &     -20.90 &      2.256 &       0.45 &         SF\\
       587734622700831054 &   08:03:29.3 &+25:44:18.6 &      0.04679 &     -21.07 &      2.295 &          - &          -\\
       587734622705811566 &   08:44:37.8 &+32:54:23.2 &      0.03154 &     -20.91 &      2.086 &          - &      LINER\\
       587734891674796073 &   11:17:59.8 &+08:44:35.0 &      0.04541 &     -20.74 &      2.255 &          - &          -\\
       587734892755419193 &   12:21:23.3 &+09:50:53.0 &      0.04667 &     -21.45 &      2.354 &          - &          -\\
       587734893287374941 &   11:35:39.9 &+10:10:02.8 &      0.04199 &     -21.39 &      2.349 &          - &          -\\
       587734949655740512 &   08:38:43.5 &+07:48:23.8 &      0.02955 &     -21.67 &      2.356 &          - &          -\\
       587735044152623112 &   09:04:55.3 &+33:57:22.1 &      0.04354 &     -21.57 &      2.376 &          - &          -\\
       587735044686348347 &   08:36:01.5 &+30:15:59.1 &      0.02561 &     -21.10 &      2.268 &          - &          -\\
       587735348575535135 &   12:23:23.8 &+15:10:18.8 &      0.04238 &     -22.04 &      2.252 &          - &     AGN+SF\\
       587735697525637156 &   14:30:51.1 &+53:54:28.0 &      0.04336 &     -21.74 &      2.348 &          - &     AGN+SF\\
       587735742615388296 &   15:53:35.6 &+32:18:20.6 &      0.04985 &     -21.07 &      1.789 &        3.2 &         SF\\
       587736477058138251 &   15:05:43.8 &+08:47:18.9 &      0.04632 &     -21.07 &      2.282 &          - &          -\\
       587736477586554950 &   13:47:47.7 &+11:16:27.0 &      0.03942 &     -21.20 &      1.911 &        5.5 &         SF\\
       587736478661279818 &   13:57:13.2 &+12:01:16.8 &      0.02077 &     -21.69 &      2.142 &          - &          -\\
       587736542015127621 &   13:52:00.4 &+08:52:55.3 &      0.03766 &     -21.48 &      1.909 &          - &     AGN+SF\\
       587736584980463705 &   16:44:30.8 &+19:56:26.7 &      0.02300 &     -20.71 &      1.873 &        5.2 &         SF\\
       587736619863965778 &   16:21:19.5 &+24:39:59.9 &      0.03786 &     -21.03 &      2.170 &          - &          -\\
       587736783604482120 &   15:47:44.4 &+37:12:18.1 &      0.03967 &     -22.24 &      2.447 &          - &          -\\
       587736941986250794 &   15:06:24.2 &+32:25:51.0 &      0.04339 &     -21.20 &      2.236 &          - &      LINER\\
       587737808499245260 &   08:37:15.4 &+55:54:29.5 &      0.03790 &     -20.80 &      2.291 &          - &          -\\
       587738067267878973 &   07:59:12.4 &+53:33:26.0 &      0.03479 &     -20.92 &      1.657 &        13.0 &         SF\\
       587738067269255432 &   08:10:20.1 &+56:12:26.3 &      0.04623 &     -20.78 &      2.130 &        1.0 &         SF\\
       587738195577929986 &   07:58:53.7 &+52:19:30.8 &      0.04070 &     -21.05 &      2.161 &          - &          -\\
       587738568174534706 &   13:34:09.4 &+13:16:51.0 &      0.04409 &     -22.03 &      2.246 &          - &     AGN+SF\\
       587738570849845253 &   12:08:37.4 &+16:08:34.0 &      0.02265 &     -21.65 &      2.401 &          - &          -\\
       587738574070939706 &   13:24:58.2 &+39:07:06.4 &      0.03726 &     -20.78 &      2.284 &          - &     AGN+SF\\
       587738618094026755 &   10:19:08.7 &+34:34:09.3 &      0.03525 &     -21.41 &      2.350 &          - &          -\\
       587738946685304841 &   12:37:15.7 &+39:28:59.3 &      0.02035 &     -20.92 &      2.156 &          - &      LINER\\
       587738947194519672 &   07:56:36.3 &+18:44:17.7 &      0.03988 &     -21.53 &      2.088 &        3.9 &         SF\\
       587739096454332452 &   12:38:52.8 &+36:32:05.7 &      0.04976 &     -20.91 &      2.173 &          - &          -\\
       587739115234394179 &   07:56:08.7 &+17:22:50.5 &      0.02899 &     -20.73 &      2.211 &        1.6 &         SF\\
       587739132429205602 &   15:57:21.4 &+24:24:28.0 &      0.04341 &     -21.60 &      2.221 &          - &     AGN+SF\\
       587739156580597805 &   10:01:01.9 &+31:12:17.1 &      0.04379 &     -21.17 &      2.091 &          - &    Seyfert\\
       587739165700653215 &   16:09:07.3 &+21:52:03.8 &      0.03127 &     -20.99 &      2.190 &          - &     AGN+SF\\
       587739166237851766 &   16:13:25.6 &+21:54:33.9 &      0.03193 &     -21.75 &      2.423 &          - &          -\\
       587739303685193778 &   13:15:37.6 &+34:02:31.5 &      0.03408 &     -20.89 &      2.218 &          - &          -\\
       587739406262468657 &   15:18:09.6 &+25:42:11.5 &      0.03260 &     -20.85 &      1.960 &        6.6 &         SF\\
       587739406268039264 &   16:07:54.0 &+20:03:03.8 &      0.03165 &     -20.73 &      1.487 &        4.8 &         SF\\
       587739506086707246 &   13:26:20.8 &+31:41:59.9 &      0.04999 &     -21.89 &      1.930 &        6.7 &         SF\\
       587739647814139970 &   10:20:34.9 &+29:14:10.8 &      0.04846 &     -20.96 &      1.998 &        1.0 &         SF\\
       587739648357826596 &   11:31:22.0 &+32:42:22.9 &      0.03368 &     -21.61 &      2.015 &        6.3 &         SF\\
       587739809414905903 &   14:30:58.8 &+22:39:45.5 &      0.04456 &     -20.79 &      2.235 &          - &          -\\
       587739810496512078 &   15:44:51.5 &+17:51:22.5 &      0.04521 &     -21.32 &      1.994 &        3.7 &         SF\\
       587739811571957865 &   16:01:28.1 &+17:14:25.9 &      0.03602 &     -21.65 &      2.427 &          - &          -\\
       587739814246023211 &   16:25:38.1 &+16:27:18.1 &      0.03432 &     -22.28 &      2.430 &          - &      LINER\\
       587739814778634376 &   15:49:59.8 &+21:11:22.7 &      0.03493 &     -20.74 &      2.184 &          - &     AGN+SF\\
       587739815852965987 &   15:56:33.7 &+21:17:20.7 &      0.01470 &     -20.95 &      2.310 &          - &          -\\
       587739828213055504 &   16:05:22.7 &+16:11:53.6 &      0.03374 &     -21.74 &      2.370 &          - &          -\\
       587739828749730189 &   16:04:39.4 &+16:44:43.6 &      0.04599 &     -20.74 &      2.308 &       0.54 &         SF\\
       587741420558221629 &   07:52:57.1 &+13:30:02.9 &      0.04926 &     -20.74 &      2.310 &          - &     AGN+SF\\
       587741490887852274 &   07:56:49.3 &+13:53:49.8 &      0.04501 &     -21.49 &      2.258 &          - &     AGN+SF\\
       587741490904105107 &   10:25:24.7 &+27:25:06.3 &      0.04973 &     -20.98 &      2.256 &        3.1 &         SF\\
       587741532763324554 &   07:55:23.3 &+12:57:11.6 &      0.04413 &     -21.29 &      2.331 &          - &     AGN+SF\\
       587741600427147349 &   13:17:39.6 &+25:32:46.3 &      0.04563 &     -21.12 &      2.118 &          - &     AGN+SF\\
       587741601491845160 &   11:45:03.7 &+26:49:39.8 &      0.03011 &     -21.24 &      2.293 &          - &    Seyfert\\
       587741602572730395 &   12:58:35.2 &+27:35:47.0 &      0.02572 &     -21.79 &      1.977 &          - &     AGN+SF\\
       587741708333940769 &   09:40:44.5 &+21:14:03.4 &      0.02442 &     -20.72 &      2.156 &          - &    Seyfert\\
       587741709410369692 &   10:05:19.9 &+23:49:01.2 &      0.04559 &     -20.74 &      2.316 &          - &     AGN+SF\\
       587741709955563564 &   11:28:19.9 &+27:37:19.6 &      0.03215 &     -20.71 &      2.320 &          - &    Seyfert\\
       587741723360165908 &   12:54:53.7 &+28:25:01.1 &      0.02465 &     -20.80 &      2.280 &          - &     AGN+SF\\
       587741816781209661 &   09:48:09.2 &+20:18:32.8 &      0.03964 &     -21.34 &      2.084 &          - &    Seyfert\\
       587741828579983420 &   10:26:39.5 &+21:48:11.6 &      0.04238 &     -21.68 &      2.387 &          - &          -\\
       587742191508586532 &   11:06:47.9 &+24:55:47.1 &      0.04851 &     -21.64 &      2.132 &          - &     AGN+SF\\
       588007005234856197 &   08:17:56.3 &+47:07:19.5 &      0.03901 &     -21.06 &      2.309 &        8.1 &         SF\\
       588007005259038814 &   14:57:46.5 &+59:20:32.7 &      0.03922 &     -21.91 &      2.335 &          - &          -\\
       588009371761573965 &   11:13:06.6 &+61:38:18.0 &      0.04543 &     -21.42 &      2.330 &          - &      LINER\\
       588010135730782222 &   09:24:29.6 &+53:41:37.8 &      0.04590 &     -20.80 &      1.997 &       0.63 &         SF\\
       588010360691490829 &   11:10:38.2 &+05:56:03.8 &      0.04147 &     -21.16 &      2.260 &          - &          -\\
       588010878765826140 &   13:01:41.4 &+04:40:49.9 &      0.03836 &     -21.40 &      2.101 &        1.5 &         SF\\
       588010879306432606 &   13:36:12.2 &+04:44:24.3 &      0.03434 &     -21.57 &      2.380 &          - &      LINER\\
       588011102644207665 &   16:02:00.2 &+47:52:54.8 &      0.04309 &     -21.27 &      2.304 &          - &     AGN+SF\\
       588011125186691149 &   12:06:17.0 &+63:38:19.0 &      0.03974 &     -21.26 &      1.686 &        18.0 &         SF\\
       588011216986505290 &   13:13:49.1 &+60:41:04.8 &      0.03809 &     -20.75 &      2.169 &          - &          -\\
       588011216991289349 &   14:32:22.7 &+56:51:08.4 &      0.04302 &     -21.74 &      1.864 &        6.1 &         SF\\
       588011218063392805 &   14:10:13.0 &+59:15:23.6 &      0.04285 &     -21.65 &      2.323 &          - &          -\\
       588013383806091298 &   09:39:14.8 &+45:21:57.3 &      0.04264 &     -20.86 &      2.310 &          - &          -\\
       588015508201472181 &   00:53:29.4 &-00:40:38.4 &      0.04377 &     -20.86 &      2.317 &          - &          -\\
       588015509815558168 &   01:25:02.9 &+00:26:39.7 &      0.02875 &     -21.59 &      2.034 &          - &     AGN+SF\\
       588016840704131407 &   07:46:02.4 &+18:43:01.7 &      0.04625 &     -21.86 &      2.420 &          - &     AGN+SF\\
       588017110752165893 &   10:46:07.3 &+44:43:28.9 &      0.04718 &     -21.95 &      2.327 &          - &     AGN+SF\\
       588017567636652100 &   12:14:15.4 &+13:30:55.8 &      0.04295 &     -20.77 &      2.311 &          - &          -\\
       588017604156784724 &   14:14:33.2 &+40:45:22.9 &      0.04185 &     -20.86 &      1.614 &        5.2 &         SF\\
       588017604680417320 &   11:29:48.7 &+44:13:49.9 &      0.04539 &     -22.13 &      2.429 &          - &      LINER\\
       588017606292865048 &   11:52:05.0 &+45:57:06.6 &      0.04316 &     -20.72 &      1.842 &        1.7 &         SF\\
       588017626159906869 &   13:17:38.0 &+43:48:38.3 &      0.02798 &     -21.07 &      1.933 &          - &     AGN+SF\\
       588017627759444009 &   10:55:46.8 &+44:02:58.8 &      0.03715 &     -21.43 &      2.270 &          - &      LINER\\
       588017627778580560 &   14:53:23.4 &+39:04:13.6 &      0.03153 &     -20.88 &      2.134 &        1.2 &         SF\\
       588017702392430629 &   11:48:57.8 &+11:27:18.4 &      0.04321 &     -20.71 &      2.053 &          - &     AGN+SF\\
       588017703476658186 &   13:26:37.5 &+11:52:10.2 &      0.04624 &     -22.49 &      2.527 &          - &          -\\
       588017704030961755 &   16:07:18.7 &+07:10:51.3 &      0.04669 &     -21.13 &      2.227 &          - &     AGN+SF\\
       588017728158695498 &   13:26:00.9 &+06:13:23.3 &      0.03961 &     -21.35 &      2.347 &          - &          -\\
       588017947211006028 &   13:37:09.3 &+39:52:24.5 &      0.04894 &     -21.51 &      2.293 &          - &          -\\
       588017978342375556 &   09:46:08.4 &+32:53:50.2 &      0.03680 &     -20.78 &      2.231 &          - &     AGN+SF\\
       588017978903232631 &   14:17:32.6 &+36:20:19.1 &      0.04712 &     -20.93 &      1.916 &        1.6 &         SF\\
       588017979974418577 &   13:50:34.6 &+38:45:06.1 &      0.04918 &     -20.74 &      2.170 &          - &          -\\
       588017979975008316 &   13:56:49.2 &+38:18:50.8 &      0.03431 &     -20.77 &      2.036 &          - &     AGN+SF\\
       588017990148816933 &   12:59:48.6 &+08:55:57.9 &      0.04620 &     -21.28 &      2.177 &          - &      LINER\\
       588017991233110227 &   14:37:33.0 &+08:04:43.0 &      0.04987 &     -21.62 &      2.353 &        3.6 &         SF\\
       588017991773978739 &   15:14:29.9 &+07:35:46.8 &      0.04484 &     -20.79 &      2.042 &          - &          -\\
       588018091080351786 &   15:58:25.4 &+32:58:07.7 &      0.04798 &     -21.87 &      2.295 &          - &          -\\
       588018254297038899 &   16:18:18.7 &+34:06:40.1 &      0.04733 &     -21.58 &      2.307 &        2.9 &         SF\\
       588023239671087134 &   09:13:20.7 &+17:38:27.7 &      0.02554 &     -21.33 &      2.238 &          - &          -\\
       588295840177061984 &   12:19:05.9 &+48:49:27.7 &      0.04468 &     -20.85 &      2.151 &          - &     AGN+SF\\
       588295842320744617 &   11:25:07.3 &+49:42:02.6 &      0.04997 &     -21.42 &      2.382 &          - &          -\\
       588297863102988421 &   08:43:46.7 &+31:34:52.6 &      0.04756 &     -20.70 &      2.110 &        7.3 &         SF\\
       588297864718909472 &   09:30:31.3 &+39:17:50.0 &      0.04605 &     -21.87 &      2.143 &          - &          -\\
       588848899365601360 &   10:26:54.6 &-00:32:29.4 &      0.03463 &     -21.38 &      2.137 &        7.6 &         SF\\
\enddata
\tablenotetext{1}{The aperture-corrected H$\alpha$ SFR. This is given only for
the objects classified as starforming by the emission line diagrams.}
\tablenotetext{2}{ Emission line classification, objects that do not have $\rm S/N >
3$ in the four main lines are given as `-'.}
\label{tab:cat}
\end{deluxetable}

\label{lastpage}

\end{document}